\documentclass[12pt,twoside,english,refpage,intoc,bibliography=totoc,index=totoc,BCOR7.5mm,captions=tableheading]{elsarticle}
\usepackage[T1]{fontenc}
\usepackage[latin9]{inputenc}
\usepackage[mathlines]{lineno}
\usepackage{color}
\usepackage{babel}
\usepackage{array}
\usepackage{float}
\usepackage{mathrsfs}
\usepackage{enumitem}
\usepackage{multirow}
\usepackage{amsmath}
\usepackage{graphicx}
\usepackage{esint}
\usepackage[unicode=true,
 bookmarks=true,bookmarksnumbered=true,bookmarksopen=false,
 breaklinks=false,pdfborder={0 0 1},backref=false,colorlinks=true]
 {hyperref}
\hypersetup{pdftitle={The LyX User's Guide},
 pdfauthor={LyX Team},
 pdfsubject={LyX},
 pdfkeywords={LyX},
 linkcolor=black, citecolor=black, urlcolor=blue, filecolor=blue, pdfpagelayout=OneColumn, pdfnewwindow=true, pdfstartview=XYZ, plainpages=false}

\makeatletter

\providecommand{\tabularnewline}{\\}

\usepackage{ifpdf} 
\ifpdf 

 \IfFileExists{lmodern.sty}{\usepackage{lmodern}}{}

\fi 

\usepackage[figure]{hypcap}

\let\myTOC\tableofcontents
\renewcommand\tableofcontents{%
  \frontmatter
  \pdfbookmark[1]{\contentsname}{}
  \myTOC
  \mainmatter }

\usepackage{fancyhdr}

\@ifundefined{extrarowheight}
 {\usepackage{array}}{}
\@ifundefined{showcaptionsetup}{}{%
 \PassOptionsToPackage{caption=false}{subfig}}
\usepackage{subfig}
\makeatother

\begin{document}

\title{Nektar++: Design and implementation of an implicit, spectral/$hp$ element, compressible flow solver using a Jacobian-free Newton Krylov approach}

\author[SKLA,aeronautics]{Zhen-Guo Yan\corref{correspondingauthor}}
\cortext[correspondingauthor]{Corresponding author}
\ead{yanzhg@mail.ustc.edu.cn}
\author[aeronautics]{Yu Pan}
\author[aeronautics]{Giacomo Castiglioni}
\author[Universit\'e-de-Li\`ege]{Koen Hillewaert}
\author[aeronautics]{Joaquim Peir\'o}
\author[exeter]{David Moxey}
\author[aeronautics]{Spencer J. Sherwin}
\address[SKLA]{State Key Laboratory of Aerodynamics, China Aerodynamics Research and Development Center, Mianyang, PRC}
\address[aeronautics]{Department of Aeronautics, Imperial College London, London, U.K.}
\address[Universit\'e-de-Li\`ege]{Universit\'e de Li\`ege, Place du 20-Ao\^ut, 7, B-4000 Li\`ege, Belgique}
\address[exeter]{Department of Mechanical Engineering, University of Exeter, Exeter, U.K.}

\begin{abstract}
	At high Reynolds numbers the use of explicit in time compressible
	flow simulations with spectral/$hp$ element discretization can become
	significantly limited by time step. To alleviate this limitation we extend
	the capability of the spectral/$hp$ element open-source software framework,
	Nektar++, to include an implicit discontinuous Galerkin compressible
	flow solver. The integration in time is carried out by a singly diagonally
	implicit Runge-Kutta method. The non-linear system arising
	from the implicit time integration is iteratively solved by the Jacobian-free Newton
	Krylov (JFNK) method. A favorable feature of the JFNK
	approach is its extensive use of the explicit operators
	available from the previous explicit in time implementation. The functionalities of different building
	blocks of the implicit solver are analyzed from the
	point of view of software design and placed in appropriate
	hierarchical levels in the C++ libraries. In the detailed implementation,
	the contributions of different parts of the solver to computational
	cost, memory consumption and programming complexity are also analyzed.
	A combination of analytical and numerical methods is adopted to simplify
	the programming complexity in forming the preconditioning matrix. The solver is verified and tested using
	cases such as manufactured compressible Poiseuille flow, Taylor-Green
	vortex, turbulent flow over a circular cylinder at $\text{Re}=3900$ and shock
	wave boundary-layer interaction. The results show that the implicit
	solver can speed-up the simulations while maintaining good simulation
	accuracy.
\end{abstract}
\begin{keyword}
	Nektar++, implicit time integration, Spectral/$hp$ element, discontinuous
	Galerkin, Jacobian-free Newton Krylov.
\end{keyword}

\maketitle

\section{Introduction}

	Nektar++ is a C++ based cross-platform open-source framework with
	the purpose of making high-order spectral/$hp$ element methods accessible
	to a wider range of researchers and engineers \citep{Cantwell2015,moxey_nektar++:_2019}.
	Similar to many packages, the Nektar++ framework is composed of libraries
	and solvers. The six core libraries provide basic functions for different
	aspects of the implementation of the spectral/$hp$ element method, including
	the elemental base functions, the extraction of geometric information,
	the numerical schemes for specific types of equations. All these libraries
	provide the basic building blocks for the numerical methods, which
	makes the construction of new solvers easier and more transparent. The Nektar++ framework
	supports various kinds of high-order curved meshes including hexahedral, prismatic, pyramidal, tetrahedral, quadrilateral
	and triangular meshes. Continuous Galerkin (CG)
	\citep{karniadakis_spectral/hp_2013}, discontinuous Galerkin (DG)
	\citep{Mengaldo2015} and flux reconstruction (FR) schemes (currently only support hexahedral and quadrilateral meshes)
	\citep{mengaldo_connections_2016} are supported for spatial discretizations.
	Up to 12 built-in solvers have
	been developed to date providing the capability of multi-solver coupling \citep{moxey_nektar++:_2019}.
	These solvers make the Nektar++ framework applicable to a wide range of simulations
	\citep{Cantwell2015,moxey_nektar++:_2019}.

	In Nektar++, the development of the implicit solver for the compressible Navier-Stokes
	(NS) equations has been based on the explicit version of the solver.
	The explicit solver has potentially significant limitations on the time
	step because of the stability restrictions of high-order spectral/$hp$
	element schemes, see for instance reference \citep{hesthaven_nodal_2007}.
	These time step restrictions
	can significantly reduce the convective and diffusive time step near
	the boundaries, e.g. to resolve boundary layers with stretched cells
	at increasing Reynolds numbers \citep{karniadakis_spectral/hp_2013}
	or near badly shaped cells in complex meshes. The unconditional
	stability of implicit solvers can not only speed-up the simulations by allowing
	much larger time steps but also reduce the risk of run time ``blow-up''
	resulting from local instability in the nonlinear evolution of the
	flow field.

	However, the development of the implicit solver presents some challenges both
	in numerical algorithms and in software design. Firstly, the coupling
	of different equations in the compressible NS equations means the
	equations cannot be solved component by component as is the case for
	existing implicit solvers of decoupled equation systems in Nektar++.
	This not only makes the coupled implicit system much larger in size
	but also makes the implicit system stiffer due to the scattering
	of eigenvalues for different equations. Secondly, the problem naturally
	leads to a nonlinear system requiring highly optimized set up operators
	since the implicit system needs to be updated along
	with the simulations.

	The development of implicit compressible flow solvers in high-order
	open-source software is still quite limited especially for unsteady simulations. Based on deal.II, Hartmann
	and Houston \citep{hartmann_symmetric_2005} developed a solver of
	the compressible NS equations using implicit time integration and
	the DG scheme. However, their solver is only tested on steady-state problems
	and only supports quadrilateral and hexahedral meshes. A fully implicit
	FR solver for compressible NS equations has been developed in the CoolFluid
	framework \citep{vandenhoeck_implicit_2019}. However, the implicit
	solver is only tested in steady state problems and only first and
	second order curved quadrilateral and hexahedral meshes are supported.
	Nek5000 provides an implicit solver based on Jacobian-free Newton
	Krylov method \citep{chudanov_validation_2014}. However, it only
	supports weakly compressible simulations through some modifications
	of the incompressible NS equations. An implicit solver based on MOOSE
	has been reported recently, but it currently only focuses on incompressible
	flow systems \citep{peterson_overview_2018}. In summary, only a few
	implicit solvers with limited capabilities are available in the high-order
	open-source community.

	We will discuss the development of an implicit solver based on the
	\mbox{Nektar++} framework. The progress and capabilities of the solver are
	summarized and demonstrated using a few test cases. The long term
	goal is to improve the efficiency and robustness of the compressible
	flow simulations in Nektar++ therefore enabling large scale high fidelity
	simulations of unsteady problems. Meanwhile, it provides a good example
	for developing user defined solvers and offers a new pattern for implicit
	solvers of coupled nonlinear systems in Nektar++. The implicit solver
	also offers a good alternative for exploring different aspects of
	the implicit solvers and for large scale simulations in the open-source
	community.

	The governing equations are presented in Section 2. Spatial and temporal
	numerical schemes are described in Section 3. Section 4 gives a detailed description of the
	software implementation and a summary of the capabilities of the implicit solver.
	Results of various test cases are presented in Section 5 to
	verify the solver implementation and to illustrate its capabilities.
	Finally, conclusions are drawn in Section 6.

\section{Governing equations}

	This section presents the Navier-Stokes governing equations of compressible
	viscous flow and discuss their non-dimensionalization.

	\subsection{Compressible Navier-Stokes equations}

		The three dimensional (3D) non-dimensional compressible NS equations can be expressed
		in abridged form as
		\begin{equation}
			\frac{\partial\mathbf{U}}{\partial t}=-\mathbf{\nabla}\cdot\mathbf{H}=-\mathbf{\nabla}\cdot(\mathbf{F}-\mathbf{G}).\label{eq:differencial-governing-equation}
		\end{equation}
		Here, the summation convention is used for repeated indexes, the conservative variable vector $\mathbf{U}$, the advection
		flux vector $\mathbf{F}=\left(\mathbf{F}_{1},\mathbf{F}_{2},\mathbf{F}_{3}\right)^{T}$
		and the diffusion flux vector $\mathbf{G}=\left(\mathbf{G}_{1},\mathbf{G}_{2},\mathbf{G}_{3}\right)^{T}$
		are
		\begin{equation}
			\mathbf{U}=\left(\begin{array}{c}
					\rho       \\
					\rho u_{1} \\
					\rho u_{2} \\
					\rho u_{3} \\
					E
				\end{array}\right),\thinspace\mathbf{F}_{i}=\left(\begin{array}{c}
					\rho u_{i}                   \\
					\rho u_{i}u_{1}+p\delta_{1i} \\
					\rho u_{i}u_{2}+p\delta_{2i} \\
					\rho u_{i}u_{3}+p\delta_{3i} \\
					u_{i}\left(E+p\right)
				\end{array}\right),\thinspace\mathbf{G}_{i}=\left(\begin{array}{c}
					0         \\
					\tau_{i1} \\
					\tau_{i2} \\
					\tau_{i3} \\
					u_{j}\tau_{ij}-q_{i}
				\end{array}\right),\label{eq:conserved-variable=000026fluxes}
		\end{equation}
		where $\delta_{ij}$ is the
		Kronecker delta, $\rho$ is the density, $p$ is the pressure, $T$
		is the temperature, $u_{i}$ is the $i$th velocity component, $E=p/\left(\gamma-1\right)+\rho u_{k}u_{k}/2$
		is the total specific energy, $\gamma=C_{p}/C_{v}=1.4$ is the ratio
		of the specific heats with constant pressure ($C_{p}$) and constant
		volume ($C_{v}$) and the stress tensor is
		\begin{equation}
			\tau_{ij}=\mu\left(\frac{\partial u_{i}}{\partial x_{j}}+\frac{\partial u_{j}}{\partial x_{i}}\right)-\frac{2}{3}\mu\frac{\partial u_{k}}{\partial x_{k}},
		\end{equation}
		where $x_{i}$ is the $i$th coordinate and $\mu$ is the dynamic viscosity.
		Finally, the $i$th component of heat flux is given by
		\begin{equation}
			q_{i}=-\frac{\mu}{\text{Ma}_{\infty}^2\text{Pr}(\gamma-1)}\frac{\partial T}{\partial x_{i}},
		\end{equation}
		where $\text{Ma}_{\infty}$ is the Mach number based on the reference flow state, $\text{Pr}$ is the Prandtl number.
		The Euler
		equations of inviscid compressible flow are recovered by ignoring the diffusion flux vector.
		Together with appropriate initial and boundary conditions, Eq. \eqref{eq:differencial-governing-equation}
		is closed through the equation of state,
		\begin{equation}
			p=\frac{\rho T}{\gamma \text{Ma}_{\infty}^2},
		\end{equation}
		and Sutherland's law,
		\begin{equation}
			\mu=\frac{T^{3/2}}{\text{Re}_{\infty}} \frac{1+110/T_{\infty}}{T+110/T_{\infty}},
		\end{equation}
		are provided. In the above equations, $\text{Re}_{\infty}$ is the Reynolds number based on the reference flow state and $T_{\infty}$ is the reference temperature which should be given in degrees Kelvin. 		

\section{Numerical methods}

	This section describes the discontinuous Galerkin method, the specific spectral/$hp$ element method adopted, and the implicit temporal discretizations.

	\subsection{Discontinuous Galerkin method}

		In the discontinuous Galerkin method, the computational domain ($\Omega$)
		is divided into $N_{e}$ non-overlapping elements ($\Omega_{e}$).
		The space of test function is defined as
		\begin{equation}
			V^{P}=\left\{ \phi:\phi|_{\Omega_{e}}\in\mathscr{P}^{P}\left(\Omega_{e}\right),e=1,\cdots,N_{e}\right\} ,\label{eq:space}
		\end{equation}
		where $\mathscr{P}^{P}\left(\Omega_{e}\right)$ is the polynomial
		space of degree $P$ in $\Omega_{e}$. The weak form of Eq. \eqref{eq:differencial-governing-equation}
		is obtained by multiplying by the test function $\phi_{p}$ and performing
		integration by parts in $\Omega_{e}$,
		\begin{equation}
			\intop_{\Omega_{e}}\frac{\partial\mathbf{U}}{\partial t}\phi_{p}d\Omega_{e}=\intop_{\Omega_{e}}\mathbf{\nabla}\phi_{p}\cdot\mathbf{H}d\Omega_{e}-\intop_{\varGamma_{e}}\phi_{p}\mathbf{H}^{n}d\varGamma_{e},\label{weak-integration}
		\end{equation}
		where $\varGamma_{e}$ is the element boundary, $\mathbf{H}^{n}=\mathbf{F}^{n}-\mathbf{G}^{n}$, $\mathbf{F}^{n}=\mathbf{F}\cdot\mathbf{n}$
		and $\mathbf{G}^{n}=\mathbf{G}\cdot\mathbf{n}$ are boundary fluxes
		on the elemental outward normal direction ($\mathbf{n}$).

		The vector of conservative variables $\mathbf{U}$ is approximated in the same
		polynomial space as the test functions and it is expressed as
		\begin{equation}
			\mathbf{U}\left(\mathbf{x},t\right)\simeq\sum_{q=1}^{N}\mathbf{u}_{q}\left(t\right)\phi_{q}\left(\mathbf{x}\right),\label{eq:coeff}
		\end{equation}
		where $\mathbf{u}_{q}\left(t\right)$ is the $q$th coefficient of
		the base (or trial) function $\phi_{q}\left(\mathbf{x}\right)$ and
		$N$ is the total degrees of freedom (DoFs) in the element. The flow
		variable values on some quadrature points $\mathbf{Q}$ are calculated
		using the backward transformation $\mathbf{Q}_{i}=\sum_{q=1}^{N}\mathbf{B}_{iq}\mathbf{u}_{q}$
		where  $\mathbf{B}$ is the backward transformation matrix with $\mathbf{B}_{iq}=\phi_{q}\left(\mathbf{x}_{i}\right)$.
		Here $\mathbf{x}_{i}$ is coordinates of the $i$th quadrature point.
		The fluxes are calculated at these quadrature points and a quadrature
		rule with $N_{Q}$ quadrature points is adopted to calculate the integration
		in the element and $N_{Q}^{\varGamma}$ quadrature points on element boundaries. This leads to
		\begin{equation}
			\begin{aligned}\sum_{i=1}^{N_{Q}}\sum_{q=1}^{N}\phi_{p}\left(\mathbf{x}_{i}\right)w_{i}J_{i}\phi_{q}\left(\mathbf{x}_{i}\right)\frac{d\mathbf{u}_{q}}{dt}= & \sum_{i=1}^{N_{Q}}w_{i}J_{i}\mathbf{\nabla}\phi_{p}\left(\mathbf{x}_{i}\right)\cdot\mathbf{H}\left(\mathbf{Q}_{i}\right) 
			\\ 
			& -\sum_{i=1}^{N_{Q}^{\varGamma}}\phi_{p}\left(\mathbf{x}_{i}^{\varGamma}\right)w_{i}^{\varGamma}J_{i}^{\varGamma}\hat{\mathbf{H}}_{i}^{n},
			\end{aligned}
			\label{eq:discrete-form-Equations}
		\end{equation}
		where $w_{i}$ and $J_{i}$ are the quadrature weights and grid metric
		Jacobian on the $i$th quadrature point, $\hat{\mathbf{H}}_{i}^{n}=\hat{\mathbf{F}}_{i}^{n}-\hat{\mathbf{G}}_{i}^{n}$, $\hat{\mathbf{F}}_{i}^{n}$
		and $\hat{\mathbf{G}}_{i}^{n}$ are the numerical normal fluxes on the
		$i$th quadrature point $\mathbf{x}_{i}^{\varGamma}$ of the element boundaries, $w_{i}^{\varGamma}$ and $J_{i}^{\varGamma}$
		are the quadrature weights and grid Jacobian of the lower dimensional integrations on element
		boundaries, which are usually not equal to the $w_{i}$ and $J_{i}$ even if the element and element boundary quadrature points are at the same position.

		The weak DG scheme for the advection term is complete as long as a
		Riemann numerical flux is used to calculate the normal flux,
		$\hat{\mathbf{F}}^{n}\left(\mathbf{Q}^{+},\mathbf{Q}^{-},\mathbf{n}\right)$,
		in which $\mathbf{Q}^{+}$ and $\mathbf{Q}^{-}$ are variable values
		on the exterior and interior sides of the element boundaries, respectively.

		\subsubsection{Interior penalty method}

			The interior penalty (IP) method is adopted to discretize the diffusion term. To simplify the complexity of the expressions for the multi-dimensional matrix operations in this section,
			all the indexes of matrices and vectors follow the summation convention.
			The diffusion flux in the IP method is
			expressed in the following equivalent form
			\begin{equation}
				\mathbf{G}_{ik}=\mathbf{K}\left(\mathbf{Q}\right)_{ijkl}\nabla_{j}\mathbf{Q}_{l},\label{eq:bi-linear-form}
			\end{equation}
			where $i$ and $j$ are indexes of different spatial directions, $k$
			and $l$ are indexes of different flow variables. The expression of
			$\mathbf{K}\left(\mathbf{Q}\right)_{ijkl}$ can be found in \citep{hartmann_symmetric_2005}.
			For clarity, only the diffusion terms are shown in the derivation.
			The IP method can be expressed in the following primal form
			\begin{equation}
				\begin{aligned}\intop_{\Omega}\frac{\partial\mathbf{U}_{k}}{\partial t}\phi_{p}d\Omega= & \sum_{e=1}^{N_{e}}\intop_{\varGamma_{e}}[\![\phi_{p}]\!]_{i}\left\{ \mathbf{K}\left(\left\{ \mathbf{Q}\right\} \right)_{ijkl}\nabla_{j}\mathbf{Q}_{l}\right\} d\varGamma_{e}-\sum_{e=1}^{N_{e}}\intop_{\Omega_{e}}\mathbf{\nabla}_{i}\phi_{p}\mathbf{G}_{ik}d\Omega_{e} 
				\\ 
				& +\beta_{s}\sum_{e=1}^{N_{e}}\intop_{\varGamma_{e}}[\![\mathbf{Q}_{l}]\!]_{i}\left\{ \mathbf{K}\left(\left\{ \mathbf{Q}\right\} \right)_{ijkl}\nabla_{j}\phi_{p}\right\} d\varGamma_{e}
				\\ 
				& +\sum_{e=1}^{N_{e}}\beta_{p}\intop_{\varGamma_{e}}[\![\phi_{p}]\!]_{i}\mathbf{K}\left(\left\{ \mathbf{Q}\right\} \right)_{ijkl}[\![\mathbf{Q}_{l}]\!]_{j}d\varGamma_{e},
				\end{aligned}
				\label{eq:IP-final}
			\end{equation}
			with $\beta_{s}=1$, $\beta_{p}=\left(P+1\right)^{2}/h$ for simulations
			with quadrilateral and hexahedral meshes, $[\![w]\!]_{i} =w^{1}\mathbf{n}_{i}^{1}+w^{2}\mathbf{n}_{i}^{2}$ and $\left\{ w\right\} =(w^{1}+w^{2})/{2}$, where the superscripts $1$ and $2$ indicate values from the two sides
			of the element interface. Compared with the IP method proposed in reference \citep{hartmann_optimal_2008},
			$\mathbf{K}\left(\mathbf{Q}\right)_{ijkl}$ is replaced by $\mathbf{K}\left(\left\{ \mathbf{Q}\right\} \right)_{ijkl}$
			for implementation purposes. $\beta_{s}=1$ makes the IP method symmetric and adjoint consistent.
			Values of $\beta_{p}$ for other mesh
			types can be found in reference \citep{hillewaert_development_2013}. 
			
			Using the relation
			between primal form and its flux form in reference \citep{arnold_unified_2002}, Eq. \eqref{eq:IP-final} can be expressed in its flux form. As a result, the whole discretization can
			be written in the following matrix form
			\begin{equation}
				\begin{aligned}\mathbf{M}\frac{d\mathbf{u}}{dt}= & \sum_{j=1}^{d}\mathbf{B}^{T}\mathbf{D}_{j}^{T}\mathbf{\Lambda}\left(wJ\right)\mathbf{H}_{j}\left(\mathbf{Q}\right)                              \\
						& -\left(\mathbf{B}^{\varGamma}\mathbf{M_{c}}\right)^{T}\mathbf{\Lambda}\left(w^{\varGamma}J^{\varGamma}\right)\hat{\mathbf{H}}^{n}
			         \\
						& -\sum_{j=1}^{d}\mathbf{B}^{T}\mathbf{D}_{j}^{T}\mathbf{J}^{T}\mathbf{\Lambda}\left(w^{\varGamma}J^{\varGamma}\right)\hat{\mathbf{S}}^{n}_{j}
				\end{aligned}
				,\label{eq:matrix-form-equations}
			\end{equation}
			where $d$ is the spatial dimension,  $\mathbf{M}=\mathbf{B}^{T}\mathbf{\Lambda}\left(wJ\right)\mathbf{B}$
			is the mass matrix, $\mathbf{\Lambda}$ represents
			a diagonal matrix, $\mathbf{D}_{j}$ is the derivative matrix in
			the $j$th direction, $\mathbf{B}^{\varGamma}$ is the backward transformation
			matrix of $\phi^{\varGamma}$ and $\mathbf{M_{c}}$
			is the mapping matrix between $\phi^{\varGamma}$ and $\phi$, $\mathbf{J}$ is the interpolation matrix from quadrature points of a element to quadrature points of its element boundaries, $\hat{\mathbf{H}}^{n}=\hat{\mathbf{F}}^{n}-\hat{\mathbf{G}}^{n}$, $\hat{\mathbf{F}}^{n}$ is the Riemann flux, $\hat{\mathbf{G}}^{n}_{k}=\mathbf{n}_{i}\left(\left\{ \mathbf{K}\left(\left\{ \mathbf{Q}\right\} \right)_{ijkl}\nabla_{j}\mathbf{Q}_{l}\right\}+\beta_{p}\mathbf{K}\left(\left\{ \mathbf{Q}\right\} \right)_{ijkl}[\![\mathbf{Q}_{l}]\!]_{j}\right)$, $\hat{\mathbf{S}}^{n}_{j,k}=\beta_{s}[\![\mathbf{Q}_{l}]\!]_{i}\left\{ \mathbf{K}\left(\left\{ \mathbf{Q}\right\} \right)_{ijkl}\right\}$. The $\hat{\mathbf{G}}^{n}_{k}$, which represents the flux integration of the first and last terms on the right-hand side of Eq. \eqref{eq:IP-final}, can be calculated together with the advection flux, while the symmetric flux $\hat{\mathbf{S}}^{n}_{j,k}$ cannot. Some approximations of the preconditioning matrices are make based on these properties of $\hat{\mathbf{G}}^{n}_{k}$ and $\hat{\mathbf{S}}^{n}_{j,k}$, as discussed in Section \ref{sec:NS-equations-related-operations}.

			Both the IP and the local DG (LDG) method are available in Nektar++ for
			the compressible flow solver. Currently the preferred implicit solver is
			based on the IP method since it is easier to derive its
			Jacobian matrix. Moreover, the IP method leads to
			smaller stencil than the LDG in general, which is beneficial not only
			for lowering the memory consumption but also for simplifying the calculation
			of the Jacobian matrix needed when preconditioning (see Section 4.2.2).

		\subsubsection{Shock-capturing method }

			An artificial viscosity method is implemented to regularize discontinuous
			solutions. This effectively amounts to adding artificial viscosity, $\mu_{av}$, and artificial thermal
			conductivity, $\kappa_{av}$, terms to the physical $\mu$ and $\kappa$ terms.
			The expressions of the artificial $\mu_{av}$ and $\kappa_{av}$ terms are
			\begin{equation}
				\mu_{av}=\mu_{0}\rho\frac{h}{P}\left(c+\sqrt{u_{k}u_{k}}\right)S,
			\end{equation}
			\begin{equation}
				\kappa_{av}=\mu_{av}\frac{C_{p}}{\text{Pr}},
			\end{equation}
			where $h$ is the mesh size, $\mu_{0}=0.25$ is a parameter that controls
			the magnitude of $\mu_{av}$ and $S$ is the shock sensor. The modal
			resolution-based shock sensor proposed in reference \citep{persson_sub-cell_2006}
			and Ducros' sensor \citep{ducros_large-eddy_1999} are implemented
			in Nektar++.

	\subsection{Time discretization}

		After the spatial discretization, the partial differential equations
		Eq. \eqref{weak-integration} become a set of ordinary differential
		equations (ODEs) of the form
		\begin{equation}
			\mathbf{M}\frac{d\mathbf{u}}{dt}=\mathbf{\mathcal{\mathcal{L}}}\left(\mathbf{u},t\right),\label{eq:ode}
		\end{equation}
		where $\mathbf{\mathcal{\mathcal{L}}}\left(\mathbf{u},t\right)$ is
		the discrete term representing the spatial discretization operator. Various time
		integration methods are implemented to discretize the temporal derivatives
		which are summarized in Section 4.3. Here we describe the explicit multi-stage
		Runge-Kutta (ERK) and the singly diagonally implicit multi-stage Runge-Kutta (SDIRK)
		methods \citep{kennedy_diagonally_2016} as illustrative examples of implementation.
		The discrete equations of the Runge-Kutta methods
		can be expressed in general as
		\begin{equation}
			\hat{\mathbf{u}}^{m}=a^{mm}\mathbf{E}^{m}+\sum_{i=0}^{m-1}a^{mi}\mathbf{E}^{i}+\mathbf{u}^{n},\thinspace m=0,2,\cdots,M-1;\label{eq:rk-stage}
		\end{equation}
		\begin{equation}
			\mathbf{E}^{m}=\triangle t\mathbf{M}^{-1}\mathbf{\mathcal{\mathcal{L}}}\left(\hat{\mathbf{u}}^{m},t^{n}+c^{m}\triangle t\right);c^{m}=\sum_{i=0}^{m}a^{mi},
		\end{equation}
		\begin{equation}
			\mathbf{u}^{n+1}=\sum_{i=0}^{M-1}b^{i}\mathbf{E}^{i}+\mathbf{u}^{n},\label{eq:rk-final}
		\end{equation}
		where $\mathbf{u}^{n}$ is the flow solution vector evaluated at the $n$th time step,
		$M$ is the total number of stages of the Runge-Kutta method, $a^{mi}$
		denotes the coefficients of the Runge-Kutta method, and $\hat{\mathbf{u}}^{m}$
		is the $m$th stage approximation of the solution. The coefficients
		of the Runge-Kutta schemes are given in Appendix A using Butcher
		tableaus.

		For explicit methods ($a^{mm}=0$), the calculation of Eq. \eqref{eq:rk-stage}
		is trivial. However, Eq. \eqref{eq:rk-stage} becomes a nonlinear
		system for implicit stages ($a^{mm}\neq0$) that can be expressed
		as
		\begin{equation}
			\begin{aligned}\mathbf{N}\left(\hat{\mathbf{u}}^{m}\right)= & \hat{\mathbf{u}}^{m}-\mathbf{S}_{m}-\triangle ta^{mm}\mathbf{M}^{-1}\mathbf{\mathcal{\mathcal{L}}}\left(\hat{\mathbf{u}}^{m},t^{n}+c^{m}\triangle t\right)=\mathbf{0}\end{aligned}
			,\label{eq:nonlinear-equation-system}
		\end{equation}
		where $\mathbf{S}_{m}=\sum_{i=0}^{m-1}a^{mi}\mathbf{E}^{i}+\mathbf{u}^{n}$
		is a source term based on known approximations at stage $m$.

		The Newton method \citep{knoll_jacobian-free_2004} is adopted for
		solving the nonlinear system \eqref{eq:nonlinear-equation-system}.
		The Newton iteration can be written as
		\begin{equation}
			\bar{\mathbf{u}}^{0}=\mathbf{S}_{m}, \label{Newton-initial-guess}
		\end{equation}
		\begin{equation}
			\frac{\partial\mathbf{N}}{\partial\mathbf{u}}\left(\bar{\mathbf{u}}^{l}\right)\left(\bar{\mathbf{u}}^{l+1}-\bar{\mathbf{u}}^{l}\right)=-\mathbf{N}\left(\bar{\mathbf{u}}^{l}\right),\thinspace l=0,1,\cdots.\label{eq:Newton-Iteration}
		\end{equation}
		When the $L_{2}$ norm of the residual is sufficiently small, namely
		\begin{equation}
			\left\Vert \mathbf{N}\left(\bar{\mathbf{u}}^{l}\right)\right\Vert _{2}<\alpha\left\Vert \mathbf{N}\left(\bar{\mathbf{u}}^{0}\right)\right\Vert _{2},\label{eq:nonlinear-critria}
		\end{equation}
		$\hat{\mathbf{u}}^{m}=\bar{\mathbf{u}}^{l}$ is regarded as the converged
		solution. A reduction of the initial residual with $\alpha=10^{-3}$
		will be enforced unless otherwise specified. In some cases, a smaller $\alpha$ is needed to maintain temporal accuracy. As discussed in Section \ref{3D_cylinder}, the solver will be slower using a smaller $\alpha$, but the change in efficiency will not be very large.  

		The restarted generalized minimal residual method (GMRES) \citep{saad_gmres:_1986}
		is used for solving the linear problem \eqref{eq:Newton-Iteration}, which is restarted when the number of GMRES iterations exceeds 30. The GMRES iteration is terminated after the residual drops by $5 \times 10^{-2}$, which is chosen mainly based on efficiency considerations. An important aspect
		in GMRES is preconditioning, which will be discussed separately. Currently, a standard version of GMRES proposed in \citep{saad_gmres:_1986} is used. Other optimized linear system solvers, such as the SNESANDERSON and SNESNGMRES \citep{brune_composing_2015} in PETSc, can also be adopted to further improve the performance of the linear system solver.

		The Jacobian matrix, $\partial\mathbf{N}/\partial\mathbf{u}$, in Eq. \eqref{eq:Newton-Iteration}
		is a large sparse matrix. The calculation and storage of $\partial\mathbf{N}/\partial\mathbf{u}$
		is usually excessively expensive compared with those of the explicit
		solver. Since $\partial\mathbf{N}/\partial\mathbf{u}$ is only used
		to calculate its inner product with a vector ($\partial\mathbf{N}/\partial\mathbf{u}\cdot\mathbf{q}$)
		in the GMRES, a Jacobian-free method is adopted to avoid explicitly
		calculating and storing the Jacobian matrix (see Section 4.2.2).

		The use of good preconditioners in GMRES is very important for efficiently
		solving stiff linear systems. Instead of solving the system of
		Eq. \eqref{eq:Newton-Iteration} directly, one can get the same solution
		by solving the preconditioned linear system
		\begin{equation}
			\left(\frac{\partial\mathbf{N}}{\partial\mathbf{u}}\mathbf{P}^{-1}\right)\left(\mathbf{P}\bigtriangleup\bar{\mathbf{u}}^{l}\right)=-\mathbf{N}\left(\bar{\mathbf{u}}^{l}\right),\label{eq:preconditioning}
		\end{equation}
		where $\mathbf{P}$ is the preconditioning matrix. A good preconditioner
		should be able to effectively cluster the distribution of eigenvalues
		of the linear system. The preconditioner $\mathbf{P}$ is usually
		an approximate matrix of the Jacobian matrix $\partial\mathbf{N}/\partial\mathbf{u}$.
		It should be relatively easy to invert since it has to be used repeatedly
		in every GMRES iteration. Usually it is not necessary to calculate
	$\partial\mathbf{N}/\partial\mathbf{u} \,\mathbf{P}^{-1}$ explicitly.
		Only the inner product of the matrix $\mathbf{P}^{-1}$
	with some vectors is required in a practical implementation.

\section{Implementation}

	This section presents the details of the software implementation of
	the implicit solver. The structure of the Nektar++ libraries and the
	general procedure followed in constructing a flow solver are introduced
	in Section 4.1. Section 4.2 describes the software design of the implicit
	solver in detail. Finally Section 4.3 summarizes the capabilities
	of the solver and the techniques for code verification.

	\subsection{Introduction to Nektar++}

		Nektar++ mainly consists of the libraries and several built-in solvers.
		The Nektar++ libraries provide  a structured hierarchy of C++ classes,
		which offers the major functions needed for spectral/$hp$ element
		methods. The
		six core libraries are:
		\begin{itemize}
			\item ${\tt LibUtilities}$: polynomial base functions, basic data classes
					such as array and matrix, linear algebra functions, time integration
					schemes, parallel communication and I/O.
			\item ${\tt StdRegions}$: standard (or reference) elements, polynomial
					approximation of the solution using base functions, operations on
					standard elements such as derivation and integration operations.
			\item ${\tt SpatialDomains}$: mapping from standard elements to physical
					elements, grid metrics and grid Jacobian.
			\item ${\tt LocalRegions}$: extension of operations on physical elements,
					physical elements inheriting from ${\tt StdRegions}$ and ${\tt SpatialDomains}$.
			\item ${\tt MultiRegions}$: collection of physical elements comprising
					$\Omega$, adjacent relations between elements, elemental and global
					variable mapping.
			\item ${\tt SolverUtils}$: high level building blocks of solvers, drivers
					controlling simulation procedures, general equation systems, specific
					numerical scheme for different types of equations.
		\end{itemize}

		In the latest version of Nektar++, four additional
		libraries have been included to enrich the libraries with functionalities
		encompassing efficient
		operator evaluation, quasi-3D simulations, high-order curved mesh
		generation and post-processing \citep{moxey_nektar++:_2019}. The
		Nektar++ libraries provide main functions related to data storage
		(such as matrix and array), linear/nonlinear algebra (such as GMRES),
		geometry related calculations and numerical methods (such as DG),
		but nothing related to specific equations.

		\begin{figure}[H]
			\begin{centering}
				\includegraphics[width=1\textwidth]{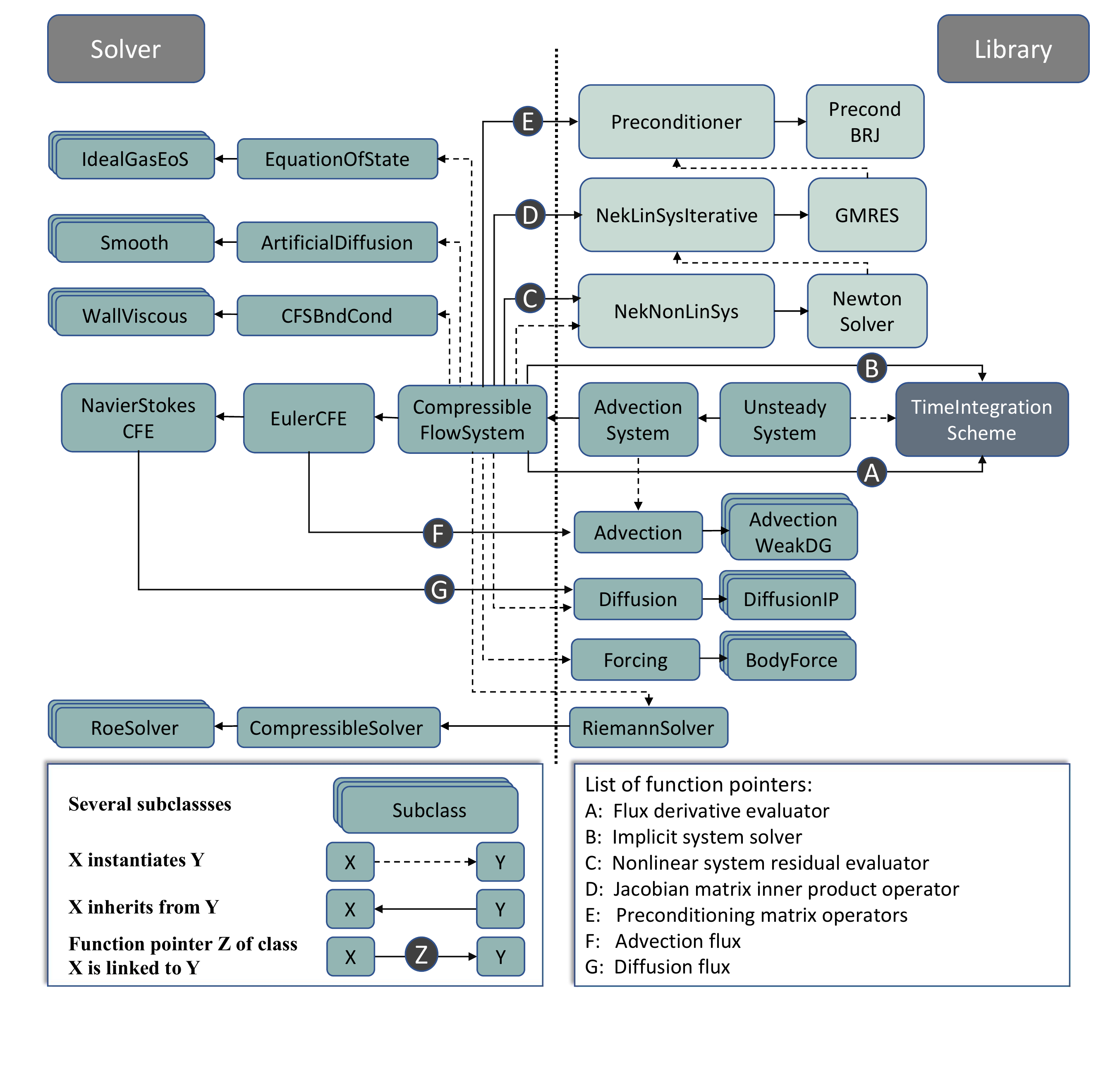}
				\par\end{centering}
			\centering{}\caption{Class structure of the explicit and implicit solvers. The
				equation system classes (${\tt EulerCFE}$ or ${\tt NavierStokesCFE}$)
				contain access to the main functionalities of the solver, such as time
				integration, and the solution fields. They make use of numerical
				methods from the libraries, such as ${\tt AdvectionWeakDG}$, and equations system related functions,
				such as the advection flux (${\tt F}$) and diffusion flux (${\tt G}$),
				to form the spatial discretization operator ${\tt A}$. The spatial
				operator ${\tt A}$ and the time integration method form the explicit
				solver using the method of lines. For the implicit solver, additional
				classes like the Newton solver (${\tt NewtonSolver}$), GMRES solver
				(${\tt GMRES}$) and linear algebra solver of preconditioners (e.g. ${\tt PrecondBRJ}$)
				are instantiated. Together with operators related to the nonlinear
				system (${\tt C}$), the Jacobian matrix (${\tt D}$) and preconditioning
				matrix (${\tt E}$), an implicit system solver (${\tt B}$) is constructed,
				which is linked to the implicit time integration scheme to form the
				implicit solver.} \label{fig:Class-structure-of}
		\end{figure}

		The main procedures for building a solver are briefly introduced using
		the explicit solver of the NS equations as an example. Besides the main
		solver class (${\tt CompressibleFlowSolver}$) which controls the
		solving procedure, an equation system class is needed. The
		equation system classes (${\tt EulerCFE}$ or ${\tt NavierStokesCFE}$)
		are instantiated and initialized dynamically based on user inputs using
		a factory method pattern \citep{Cantwell2015}, which is also extensively
		used for the dynamic object creation of classes in the solver. The equation
		system class inherits instantiations of classes related to geometry
		information, solution approximation, time integration and others from
		equations system classes in the libraries such as ${\tt UnsteadySystem}$
		in ${\tt SolverUtils}$. Thus the main functionality of the equation system class
		is to offer equation system related functions and to form spatial
		discretization operators using numerical methods from the libraries
		such as, for instance, ${\tt AdvectionWeakDG}$. Fig. \ref{fig:Class-structure-of}
		illustrates the class structure of the explicit and implicit solvers.
		The equation of state (${\tt EquationOfState}$), boundary conditions
		(${\tt CFSBndCond}$), Riemann flux (${\tt RiemmanSolver}$), shock
		capturing method (${\tt ArtificialDiffusion}$), forcing term (${\tt Forcing}$),
		advection flux (function pointer ${\tt F}$) and diffusion flux (function
		pointer ${\tt G}$) are instantiated or implemented in the equation system class.
		Specific numerical schemes like the weak
		DG scheme for advection terms (${\tt AdvectionWeakDG}$) and the interior
		penalty method for diffusion terms (${\tt DiffusionIP}$) are instantiated
		to calculate $\mathbf{\nabla}\cdot\mathbf{F}$
		and $\mathbf{\nabla}\cdot\mathbf{G}$ in Eq. \eqref{eq:differencial-governing-equation}
		provided with all these equation system related functions (${\tt F}$ and ${\tt G}$). Finally
		the spatial discretization operator $\mathbf{\mathcal{\mathcal{L}}}\left(\mathbf{u},t\right)$
		is linked to the time integration class using the pointer-to-function
		${\tt A}$ and the explicit solver is complete.

	\subsection{Implicit solver architecture and implementation}

		Section 4.2.1 presents the analysis of different building blocks of
		the implicit solver and their hierarchical structures. The implementation
		of the operations related to the NS equations is then discussed in
		Section 4.2.2.

		\subsubsection{Analysis of the implicit solver architecture}

			We first analyze the different building blocks of the solver from the
			point of view of library and software design and place
			these blocks into the right hierarchical levels in the libraries and
			the solver. 
			The Newton method, GMRES method and linear
			algebra solvers of preconditioners, which are general mathematical methods
			independent of the type of equations, are placed
			in the ${\tt LibUtilities}$ as shown on the right upper corner of Fig.~
			\ref{fig:Class-structure-of}. Meanwhile operations that depend on the equations and the numerical schemes, i.e. the implicit system solver
			(function pointer ${\tt B}$ in Fig.~\ref{fig:Class-structure-of}),
			nonlinear system residual evaluator (function pointer ${\tt C}$ in
			Fig.~\ref{fig:Class-structure-of}), Jacobian matrix operation (function
			pointer D in Fig.~\ref{fig:Class-structure-of}) and preconditioning
			matrix operations (function pointer ${\tt E}$ in Fig.~\ref{fig:Class-structure-of}),
			are coded at the solver level in the ${\tt CompressibleFlowSystem}$
			class and they are linked to the corresponding classes from the libraries
			using pointers to these functions.

			This structure provides a general framework for developing implicit
			solvers for different equation systems. Provided with equation system
			related functions, an implicit solver can also be constructed for
			other equations systems using the classes programmed in the libraries.

		\subsubsection{Implementation of the NS equations related operations}\label{sec:NS-equations-related-operations}

			The top level class
			structure of the implicit solver is illustrated in Fig.~\ref{fig:Class-structure-of}.
			Even though only a nonlinear system
			solver is added to the Nektar++ libraries, the nonlinearity and coupling
			of the NS equations makes the design of the Jacobian matrix and preconditioning
			matrix related operations much more difficult than that of the existing
			implicit solvers of linear decoupled equation systems for the following reasons: a) the nonlinearity means the Jacobian and preconditioning matrices should be updated along with the simulation, which makes the
			computational speed of these operations vital for the overall
			efficiency; b) The coupling of eigenvalues of different equations
			increases the stiffness of the system; c) The coupling of different equations leads to a much larger Jacobian matrix, especially in large-scale 3D simulations. We will discuss
			the design of equation related operations in the following.

			\paragraph{Nonlinear system residual}

			The nonlinear system residual, $\mathbf{N}(\mathbf{u})$, is easily
			available from an existing explicit solver. However, for the sake of efficiency it is important
			to optimize its evaluation since the implicit solver requires repeated calculations
			of the residual. Techniques such as sum-factorization
			\citep{cantwell_h_2011,orszag_spectral_1980} and collections of similar linear algebra
			operations \citep{moxey_nektar++:_2019}, have been studied and adopted
			in Nektar++ to speed-up the evaluation of $\mathbf{N}(\mathbf{u})$.

			\paragraph{Jacobian matrix calculation}

			Explicitly calculating and storing the Jacobian matrix $\partial\mathbf{N}/\partial\mathbf{u}$
			accurately is expensive both in CPU cost and memory consumption, as discussed in Section \ref{sec:Performance-memory}. In the GMRES solver, we do not need to calculate the Jacobian matrix explicitly, but
			only need to evaluate the inner product of the Jacobian matrix with a vector
			($\partial\mathbf{N}/\partial\mathbf{u}\cdot\mathbf{q}$), therefore
				we adopt a Jacobian-free method which reads
				\begin{equation}
					\frac{\partial\mathbf{N}}{\partial\mathbf{u}}\left(\mathbf{u}\right)\cdot\mathbf{q} \simeq \frac{\mathbf{N}(\mathbf{u}+\epsilon\mathbf{q})-\mathbf{N}(\mathbf{u})}{\epsilon},\label{eq:Jacobian-free-matrix}
				\end{equation}
				\begin{equation}
					\epsilon=\left\Vert \mathbf{u}\right\Vert _{2}\sqrt{\hat{\epsilon}},\label{eq:eps-JF}
				\end{equation}
			where $\hat{\epsilon}=2.22\times10^{-16}$ is the machine epsilon 
			for our computation system. By storing $\mathbf{N}(\mathbf{u})$,
			only one evaluation of the nonlinear residual $\mathbf{N}\left(\mathbf{u}+\epsilon\mathbf{q}\right)$
			is required for each $\mathbf{q}$. The Jacobian-free method
			offers a relatively efficient way of calculating the Jacobian matrix, while keeping enough accuracy to maintain high convergence speed \citep{knoll_jacobian-free_2004}. The errors in numerically calculating the Jacobian matrix have been studied in references \citep{vanden_comparison_1996,ezertas_performances_2009}. Both references report similar convergence history using numerical Jacobian and analytical Jacobians. However, reference \citep{xiaoquan_robust_2019} finds that simulations with analytical Jacobian has better efficiency.

			From the point of view of software design, this replaces a numerical method and a set of operations
			that are highly dependent of the type of equation solved by a more general finite
			difference scheme and an existing residual evaluator $\mathbf{N}(\mathbf{u})$.
			The Jacobian-free code will also be generally applicable to other
			equations systems if provided with the corresponding $\mathbf{N}(\mathbf{u})$.

			\paragraph{Preconditioning matrix calculation}

			Constructing a good preconditioning matrix $\mathbf{P}$ is highly
			dependent on appropriately approximating the
			Jacobian matrix $\partial\mathbf{N}/\partial\mathbf{u}$.
			Although some explorations on completely matrix-free preconditioners
			have been reported in the literature, the most efficient and most
			widely used preconditioners still require explicitly evaluating at
			least part of $\partial\mathbf{N}/\partial\mathbf{u}$ for low to medium order simulations \citep{Franciolini2017,knoll_jacobian-free_2004}.
			Here a $J$ step block relaxed Jacobi iterative
			preconditioner, BRJ($J$), is developed instead
			of the widely adopted incomplete LU factorization method (ILU) mostly
			due to its lower memory requirements.

			The whole Jacobian matrix $\partial\mathbf{N}/\partial\mathbf{u}$
			can be divided into $N_{e}\times N_{e}$ small blocks, where $N_{e}$
			is the number of elements. The $e_{1}$th row and $e_{2}$th column
			block is $\partial\mathbf{N}_{e_{1}}/\partial\mathbf{u}_{e_{2}}$
			where $\mathbf{N}_{e_{1}}$ is the nonlinear system residual of the
		$e_{1}$th element and $\mathbf{u}_{e_{2}}$ is the independent variables
			of the $e_{2}$th element. The matrix $\partial\mathbf{N}/\partial\mathbf{u}$
			can be decoupled into
			\begin{equation}
				\frac{\partial\mathbf{N}}{\partial\mathbf{u}}=\mathbf{\hat{L}}+\hat{\mathbf{D}}+\hat{\mathbf{U}},
					\label{eq:L+D+U}
			\end{equation}
			where $\mathbf{\hat{L}}$, $\hat{\mathbf{D}}$ and $\hat{\mathbf{U}}$
			correspond to the lower ($e_{1}<e_{2}$), diagonal ($e_{1}=e_{2}$)
			and upper ($e_{1}>e_{2}$) block part of $\partial\mathbf{N}/\partial\mathbf{u}$,
			respectively. Then the product $\mathbf{P}^{-1}\mathbf{q}$ is implicitly
			calculated through the iteration,
			\begin{equation}
				\begin{aligned}\hat{\mathbf{q}}^{0} & =\mathbf{0},                                                                                                                                                                                        \\
					\hat{\mathbf{q}}^{j} & =\omega\hat{\mathbf{D}}^{-1}\left[\mathbf{q}-\left(\mathbf{\hat{L}}+\hat{\mathbf{U}}\right)\hat{\mathbf{q}}^{j-1}\right]+\left(1-\omega\right)\hat{\mathbf{q}}^{j-1},\thinspace j=1,2,\cdots,J,
				\end{aligned}
				\label{eq:precond-Jacobi}
			\end{equation}
			and $\hat{\mathbf{q}}^{J}$ is the preconditioned vector. Currently
			the case with $\omega=1$ is studied. To minimize
			the memory consumption, only the matrix $\text{\ensuremath{\hat{\mathbf{D}}^{-1}}}$
			is stored while the product $\left(\mathbf{\hat{L}}+\hat{\mathbf{U}}\right)\hat{\mathbf{q}}^{j-1}$
			is calculated on the fly. $J=7$ is used in this paper, if not specified.

			The matrix $\partial\mathbf{N}_{e_{1}}/\partial\mathbf{u}_{e_{2}}$ can be derived from Eqs \eqref{eq:matrix-form-equations} and \eqref{eq:nonlinear-equation-system}
			as
			\begin{equation}
				\begin{aligned}\frac{\partial\mathbf{N}_{e_{1}}}{\partial\mathbf{u}_{e_{2}}}= & \mathbf{I}\delta_{e_{1}e_{2}}-\triangle ta^{mm}\mathbf{M}^{-1}\frac{\partial\mathcal{\mathcal{L}}_{e_{1}}}{\partial\mathbf{u}_{e_{2}}}\end{aligned}
				,\label{eq:Jacobian-matrix-form}
			\end{equation}
			\begin{equation}
				\begin{aligned}\frac{\partial\mathcal{\mathcal{L}}_{e_{1}}}{\partial\mathbf{u}_{e_{2}}}= & \sum_{j=1}^{d}\mathbf{B}_{e_{1}}^{T}\mathbf{D}_{j,e_{1}}^{T}\mathbf{\Lambda}\left[w_{e_{1}}J_{e_{1}}\left(\frac{\partial\mathbf{H}_{j,e_{1}}}{\partial\mathbf{Q}_{e_{1}}}\right)_{\nabla\mathbf{Q}}\right]\mathbf{B}_{e_{1}}\delta_{e_{1}e_{2}}+                                                \\
				& \sum_{i=1}^{d}\sum_{j=1}^{d}\mathbf{B}_{e_{1}}^{T}\mathbf{D}_{j,e_{1}}^{T}\mathbf{\Lambda}\left[w_{e_{1}}J_{e_{1}}\left(\frac{\partial\mathbf{H}_{j,e_{1}}}{\partial\nabla_{i}\mathbf{Q}_{e_{1}}}\right)_{\mathbf{Q}}\right]\mathbf{D}_{i,e_{1}}\mathbf{B}_{e_{1}}\delta_{e_{1}e_{2}}+                                                \\
				& \left(\mathbf{B}_{e_{1}}^{\varGamma}\mathbf{M}_{c}\right)^{T}\mathbf{\Lambda}\left[w^{\varGamma}J^{\varGamma}\left(\frac{\partial\hat{\boldsymbol{\mathbf{H}}}_{e_{1}}^{n}}{\partial\hat{\mathbf{Q}}^{+/-}}\right)_{\nabla\mathbf{Q}}\right]\frac{\partial\hat{\mathbf{Q}}^{+/-}}{\partial\mathbf{Q}_{e_{2}}}\mathbf{B}_{e_{2}} +
				\\
				& \sum_{j=1}^{d}\left(\mathbf{B}_{e_{1}}^{\varGamma}\mathbf{M}_{c}\right)^{T}\mathbf{\Lambda}\left[w^{\varGamma}J^{\varGamma}\left(\frac{\partial\hat{\boldsymbol{\mathbf{H}}}_{e_{1}}^{n}}{\partial\nabla_{j}\hat{\mathbf{Q}}^{+/-}}\right)_{\mathbf{Q}}\right]\frac{\partial\nabla_{j}\hat{\mathbf{Q}}^{+/-}}{\partial\nabla_{j}\mathbf{Q}_{e_{2}}}\mathbf{D}_{j,e_{2}}\mathbf{B}_{e_{2}}
				\\
				& + \sum_{j=1}^{d}\mathbf{B}^{T}\mathbf{D}_{j}^{T}\mathbf{J}^{T}\mathbf{\Lambda}\left[w^{\varGamma}J^{\varGamma}\left(\frac{\partial\hat{\boldsymbol{\mathbf{S}}}_{e_{1}}^{n}}{\partial\hat{\mathbf{Q}}^{+/-}}\right)\right]\frac{\partial\hat{\mathbf{Q}}^{+/-}}{\partial\mathbf{Q}_{e_{2}}}\mathbf{B}_{e_{2}},
				\end{aligned}
				\label{eq:Jacobian-matrix-form2}
			\end{equation}
			where $\partial\hat{\mathbf{Q}}^{+/-}/\partial\mathbf{Q}_{e_{2}}$
			is the interpolation matrix from $\mathbf{Q}_{e_{2}}$ to $\hat{\mathbf{Q}}^{+/-}$
			which is independent of $\mathbf{Q}$ itself, superscript $+/-$ is used to indicate that only one of them is a non-zero matrix for a specific $e_{2}$. The term $\partial\nabla_{j}\hat{\mathbf{Q}}^{+/-}/\partial\nabla_{j}\mathbf{Q}_{e_{2}}$ is similarly an interpolation matrix.

			The first two terms on the right-hand side of Eq. \eqref{eq:Jacobian-matrix-form2} are element integration evaluations, which are calculated using analytical methods. Following the ideas described in reference \citep{hillewaert_development_2013}, single precision data type, continuous memory operations and matrix padding are used to reduce the CPU cost and optimize cache performance in the implementation of these evaluations. 
			
			The remaining three terms are element boundary terms. Approximations are adopted to reduce the cost of preconditioner and simplify the coding. The last two terms on the right-hand side of Eq. \eqref{eq:Jacobian-matrix-form2} are neglected in all the simulations for the following two reasons. First, they are much more expensive than the third term on the right-hand side of Eq. \eqref{eq:Jacobian-matrix-form2}, which has similar computation cost as the boundary integration of the advection flux only. Considering the diffusion term is much more expensive in 3D NS simulations, this simplification makes the computation cost of  $\left(\mathbf{\hat{L}}+\hat{\mathbf{U}}\right)\hat{\mathbf{q}}^{j-1}$ much smaller than one $\mathbf{N}(\mathbf{u})$ evaluation, as discussed in Section \ref{sec:Performance-analysis}. Second, numerical tests show that neglecting these two terms has very small impact on the preconditioning performance. For instance, neglecting the last two terms in the compressible Poiseuille flow test case in Section \ref{sec:accuracy-test} with $P=2$ and $20^2$ meshes only increases the total number of GMRES iterations in the first 10 time steps from 1136 to 1144.
			
			To further simplify the coding complexity, the flux Jacobians $\partial\hat{\boldsymbol{\mathbf{H}}}_{e_{1}}^{n}/\partial\hat{\mathbf{Q}}^{\pm}$, the only functions of $\mathbf{Q}$ in the third term on the right-hand side of Eq. \eqref{eq:Jacobian-matrix-form2}, are calculated using a finite difference approximation to avoid coding the Jacobian matrices of various Riemann fluxes and boundary conditions.
			The finite difference approximation adopted is
			\begin{equation}
				\left(\frac{\partial\hat{\boldsymbol{\mathbf{H}}}_{e_{1}}^{n}}{\partial\hat{\mathbf{Q}}^{\pm}}\right)_{\nabla\mathbf{Q}} \simeq \frac{\hat{\boldsymbol{\mathbf{H}}}_{e_{1}}^{n}\left(\hat{\mathbf{Q}}^{\pm}+\chi\mathbf{e}_{j},\hat{\mathbf{Q}}^{\mp}, \nabla\hat{\mathbf{Q}}\right)-\hat{\boldsymbol{\mathbf{H}}}_{e_{1}}^{n}\left(\hat{\mathbf{Q}}^{\pm},\hat{\mathbf{Q}}^{\mp}, \nabla\hat{\mathbf{Q}}\right)}{\chi},\label{eq:numerical-Jac-on-traces}
			\end{equation}
			where $\mathbf{e}_{j}$ is the vector with one on the $j$th column
			but zeros on others. The parameter $\chi$ is evaluated
			in a fashion similar to $\epsilon$ in Eq. \eqref{eq:eps-JF}. However, the $L_2$ norm $\left\Vert \mathbf{u}\right\Vert _{2}$ in Eq. \eqref{eq:eps-JF} is replaced by
			the $L_2$ norm of each variable, which makes $\chi$ different for each variable.
			Specific physical boundary
			conditions are imposed on physical boundaries
			in the $\hat{\boldsymbol{\mathbf{H}}}_{e_{1}}^{n}$
			evaluation so that the Jacobian matrices of the boundary conditions
			are already included in Eq. \eqref{eq:numerical-Jac-on-traces}. Because
			of the independence of $\hat{\boldsymbol{\mathbf{H}}}^{n}$ on each
			quadrature point, in total $2N_{var}$ numerical flux evaluations on all the quadrature
			points of element boundaries and $N_{var}$ boundary treatment operations are required to calculate $\partial\hat{\boldsymbol{\mathbf{H}}}_{e_{1}}^{n}/\partial\hat{\mathbf{Q}}^{\pm}$.
			The matrices of $\partial\hat{\boldsymbol{\mathbf{H}}}_{e_{1}}^{n}/\partial\hat{\mathbf{Q}}^{\pm}$
			are stored for the evaluation of $\left(\mathbf{\hat{L}}+\hat{\mathbf{U}}\right)\hat{\mathbf{q}}^{l-1}$
			in Eqs. \eqref{eq:precond-Jacobi}. The computation cost and storage
			consumption of the matrix $\partial\hat{\boldsymbol{\mathbf{H}}}_{e_{1}}^{n}/\partial\hat{\mathbf{Q}}^{\pm}$
			is small compared with the other parts of the implicit solver, as shown in Section \ref{sec:Performance-memory}.

			The hybrid analytical and numerical method offers an efficient and
			accurate way of calculating the preconditioning matrices.
			In addition, it greatly reduces the complexity in programming. For simulations with shocks,
			the terms with $\partial\mu_{av}/\partial\mathbf{u}$
			and $\partial\kappa_{av}/\partial\mathbf{u}$ are neglected in the preconditioning matrices
			calculations. As shown in Section \ref{SWBLI}, the implicit solver is able to stably and efficiently simulate discontinuous flows neglecting these two terms. The preconditioning matrices are frozen for 10 time steps
			in the following simulations, if not specified.

\begin{figure}[H]
				\begin{centering}
					\includegraphics[width=1\textwidth]{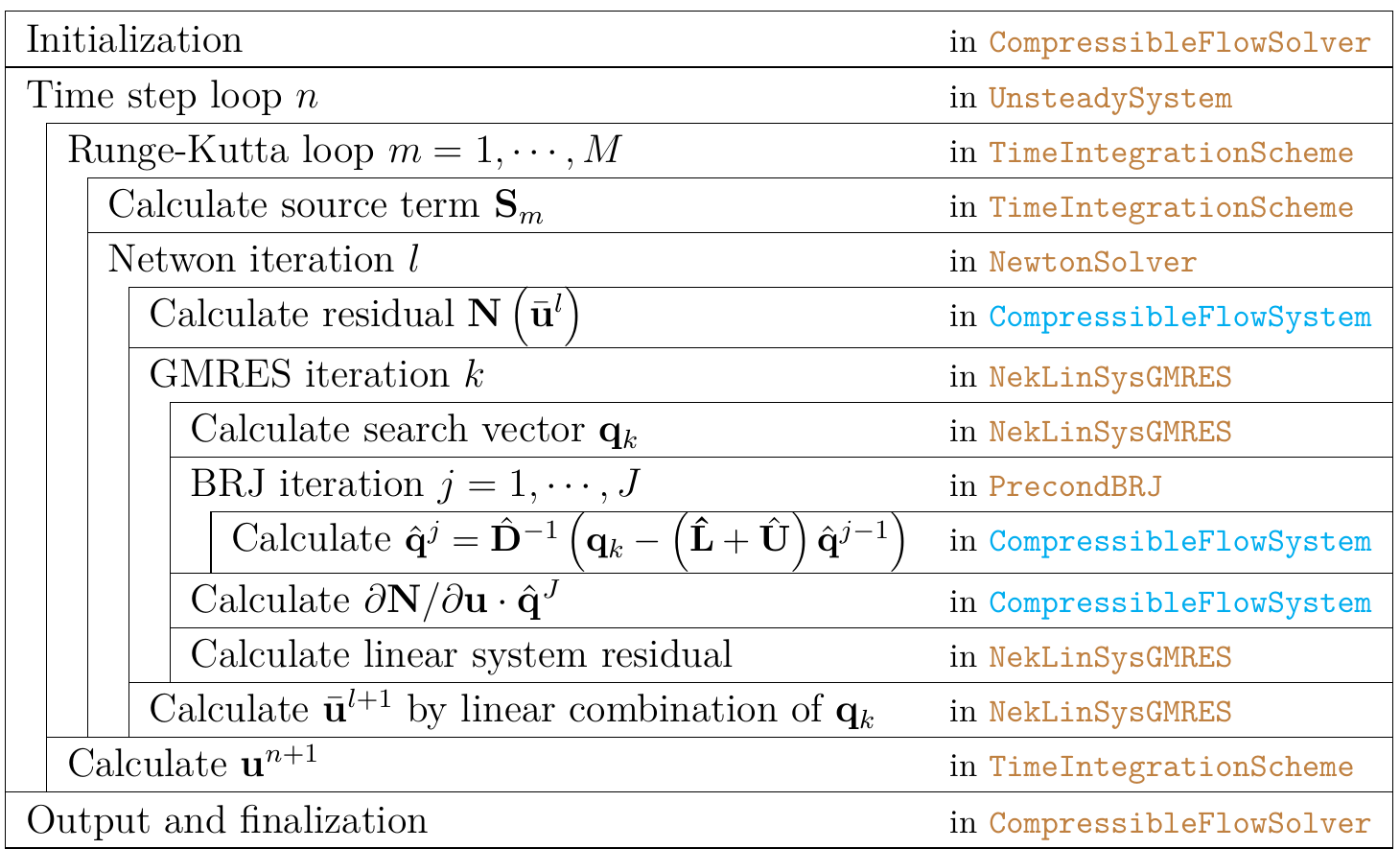}
					\par\end{centering}
				\centering{}\caption{Nassi--Shneiderman diagram of the implicit solver with the corresponding
					class names. Brown: library class ; cyan: solver class. \label{fig:Flow-chart-of}}
			\end{figure}

			The main simulation flowchart of the implicit solver is given in Fig. \ref{fig:Flow-chart-of},
			which also shows the flow between different classes.

	\subsection{Solver capabilities and code verification}

		The main capabilities of the explicit and implicit compressible flow
		solvers are summarized here. The explicit solver can use forward Euler,
		Adams Bashforth \citep{Vos2011} and 2nd-4th order Runge-Kutta for
		temporal discretization, while the implicit solver can use backward
		Euler, 2nd order BDF (Backward differentiation formula) and 2nd-4th
		order SDIRK. The other aspects of the solvers are summarized as follows:
		\begin{itemize}
			\item Supported equation systems: Euler, Navier-Stokes;
			\item Advection  discretization schemes: DG (with various nodal or modal
					bases functions), FR \citep{mengaldo_connections_2016} (not supported
					by the implicit solver yet);
			\item Diffusion discretization schemes: IP \citep{hartmann_symmetric_2005}
					(can recover a specific version of direct DG \citep{cheng_direct_2016}), LDG \citep{cockburn_local_1998}
					(the implicit solver is partially supported);
			\item Riemann numerical flux: AUSM, Roe, HLL, HLLC, Steger-Warming,
					Lax-Friedrichs and and the iterative exact Riemann solver in Section 4.9 of \citep{toro_riemann_2009};
			\item Boundary conditions: periodic, time dependent Dirichlet, symmetry,
					slipping wall, isothermal non-slip wall, adiabatic non-slip wall, Riemann invariant,
					pressure inflow, pressure outflow, zero-order extrapolation \citep{Mengaldo2015};
			\item Shock capturing: modal resolution-based shock sensor \citep{persson_sub-cell_2006},
					Ducros' sensor \citep{ducros_large-eddy_1999}, physical and Laplace-based artificial
					viscosity;
			\item Mesh types: triangular, quadrilateral, hexahedral, prismatic, pyramidal,
					tetrahedral meshes;
			\item Parallel: MPI based parallelization for HPC computing.
		\end{itemize}
		All of these options are supported by the explicit solver, but are
		only partially supported by the implicit one. Options not supported
		or partially supported by the implicit solver have been marked above.

		CTest facility of the CMake environment \citep{Cantwell2015} are adopted to execute hundreds of test cases for testing different 	building blocks of the solver. 
		These tests are automatically executed on a variety of platforms using
		a buildbot service to ensure the reliability of the solver \citep{Cantwell2015}.

\section{Performance and memory consumption of the implicit solver} \label{sec:Performance-memory}

	The implicit solver can significantly relax the constraint on the time step. However, it also requires much larger computational cost and memory consumption. The computational performance of different parts of the implicit solver is analyzed in Section \ref{sec:Performance-analysis}, and the memory consumption is discussed in Section \ref{sec:Memory-analysis}.

	\subsection{Performance analysis}\label{sec:Performance-analysis}
		
		In order to provide a deeper insight of the implicit solver implementation, the computational costs of the different parts of the implicit solver are evaluated in the following. The evaluations are performed on a CentOS 7.7.1908 using standard -O3 compiler optimizations with the gcc 5.4.0 C++ compiler and version 2019.5.281 of the BLAS and Lapack implementation in the Intel MKL library. The evaluations are run on the HPC cluster of Imperial College London with one node and four nodes for the 2D and 3D cases, respectively. Each node is equipped with two Intel Xeon E5-2620 CPUs and 120 GB RAM. The computation CPU time is obtained using Oracle Developer Studio 12.6.

		\begin{figure}[H]
			\begin{centering}
				\subfloat[The GMRES and preconditioning matrix calculations in the implicit solver \label{fig:PerfAnaly-3D-whole}]{\begin{centering}
						\includegraphics[width=0.5\textwidth]{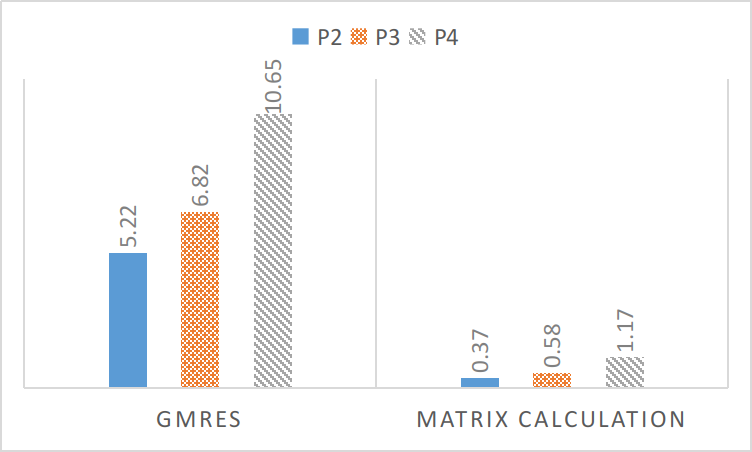}
						\par\end{centering}
					\begin{centering}
						\par\end{centering}
				}
				
				\subfloat[Operations in GMRES \label{fig:PerfAnaly-3D-GMRES}]{\begin{centering}
						\includegraphics[width=0.5\textwidth]{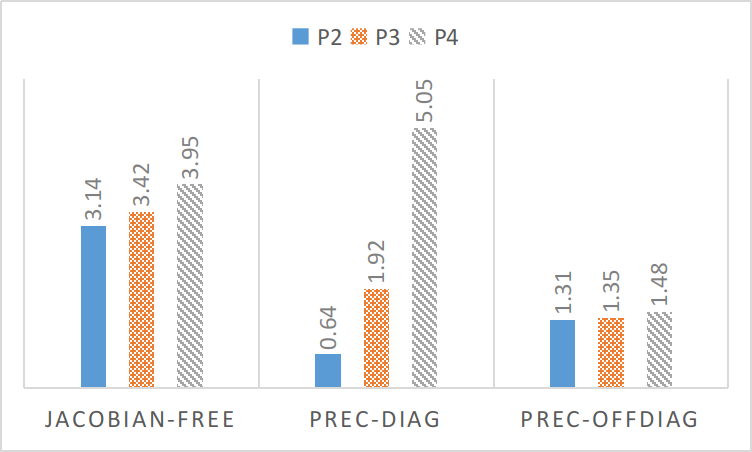}
						\par\end{centering}
					\begin{centering}
						\par\end{centering}
				}\subfloat[Operations in preconditioning matrices calculation\label{fig:PerfAnaly-3D-Prec}]{\begin{centering}
						\includegraphics[width=0.5\textwidth]{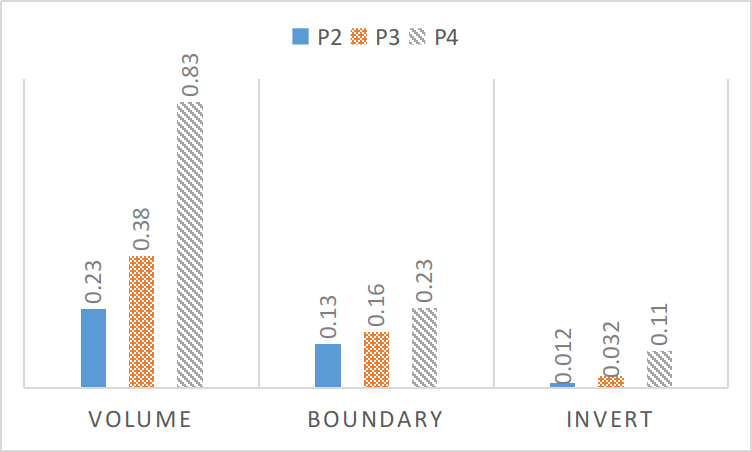}
						\par\end{centering}
					\begin{centering}
						\par\end{centering}
				}
				\par\end{centering}
			\begin{centering}
				\caption{Performance analysis of the implicit solver in 3D \label{fig:PerfAnaly-3D}}
				\par\end{centering}
		\end{figure}

		\subsubsection{3D case} \label{performance-analy-3D}
			The 3D cylinder test case described  in Section \ref{3D_cylinder} is adopted here for the performance analysis. 
			The simulations are carried out at polynomial orders $P=2$, $P=3$ and $P=4$, using seven preconditioning iterations ($J=7$) and a time step corresponding to CFL=40. The Newton iteration tolerance is $\alpha=10^{-3}$. In total 10 steps are simulated, with the preconditioning matrices calculated at the beginning of the simulations and frozen in the following 10 time steps.  Fig. \ref{fig:PerfAnaly-3D} shows the CPU time spent in different parts of the implicit solver, which is normalized by the total CPU time spent in evaluating $\mathbf{N}(\mathbf{u})$ the same number of times as the total number of Newton iterations. Fig. \ref{fig:PerfAnaly-3D-whole} shows that the GMRES procedure, which calls the Jacobian-free operation and the preconditioners, makes up the majority of CPU time. The calculation of preconditioning matrices is much cheaper because of the freezing of the matrices. However, if the preconditioning matrices are updated every time step, their computational cost will be similar to that of the GMRES procedure and the implicit solver will be about twice as expensive as the current simulations.

			Fig. \ref{fig:PerfAnaly-3D-GMRES} shows the three most expensive operations in the GMRES procedure: the JACOBIAN-FREE, PREC-OFFDIAG and PREC-DIAG. The JACOBIAN-FREE procedure corresponds to the operation of Eq. \eqref{eq:Jacobian-free-matrix}, in which the majority of CPU time is spent on one $\mathbf{N}(\mathbf{u})$ evaluation. As a result, the value of the normalized CPU time of the JACOBIAN-FREE correctly indicates the averaged number of GMRES iterations in each Newton iteration. The PREC-OFFDIAG corresponds to the $(\hat{\mathbf{L}}+\hat{\mathbf{U}})$ vector inner product in Eq.~\eqref{eq:precond-Jacobi}, while the PREC-DIAG corresponds to  the $\hat{\mathbf{D}}^{-1}$ vector inner product in Eq.~\eqref{eq:precond-Jacobi}. Since the BRJ(7) is used with $\mathbf{0}$ initial guess, six PREC-OFFDIAG and seven PREC-DIAG are needed for each JACOBIAN-FREE operation. Although much more preconditioning operations are performed, their costs are smaller than that of the JACOBIAN-FREE operations for $P=2$ and $P=3$. Even though we have already adopted the single precision data type to reduce the operations of the matrix-vector inner product, the $\hat{\mathbf{D}}^{-1}$ vector inner product still grows much faster than the other two parts when $P$ increases. This is mainly because it uses a matrix-based implementation while most operations in the other parts use a sum-factorization implementation. This also indicates that matrix-based preconditioners will easily lose efficiency for high-order 3D simulations. Techniques proposed in reference \citep{bastian_matrix-free_2019} may be used to implement a matrix-free version of the BRJ preconditioner to avoid the fast growth of the $\hat{\mathbf{D}}^{-1}$ vector inner product. As for the off-diagonal operations, the simplifications and implementations make sure they are relatively cheap compared with other parts.

			Fig. \ref{fig:PerfAnaly-3D-Prec} shows the most expensive parts in the preconditioning matrices calculations. For the element volume part, techniques such as using single precision, maintaining memory continuous and padding the matrices as discussed in reference \citep{hillewaert_development_2013}, have been adopted in our implementation to speed-up the calculation. However, this is still the most expensive part. In our implementation, the most expensive part of the volume term is implemented using BLAS::sger, but it still grows quite fast as $P$ increases. Inverting $\hat{\mathbf{D}}$, which is implemented using Lapack::dgetrf and Lapack::dgetri, only makes up a small proportion in the test cases. However, it grows the fastest, which indicates it will become the dominant part when the polynomial order is high enough. In this case, techniques such as approximate matrix inversion \citep{bastian_matrix-free_2019,pazner_approximate_2018,diosady_scalable_2019} may be adopted.

			Fig.~\ref{fig:PerfAnaly-3D} shows that a large proportion of computation costs in the implicit solver is from operations shared with the explicit solver. Therefore, the implicit solver will also greatly benefit from improvements in efficiency of the explicit solver.

			\begin{figure}[H]
				\begin{centering}
					\subfloat[The GMRES and preconditioning matrix calculations in the implicit solver \label{fig:PerfAnaly-3D-whole}]{\begin{centering}
							\includegraphics[width=0.5\textwidth]{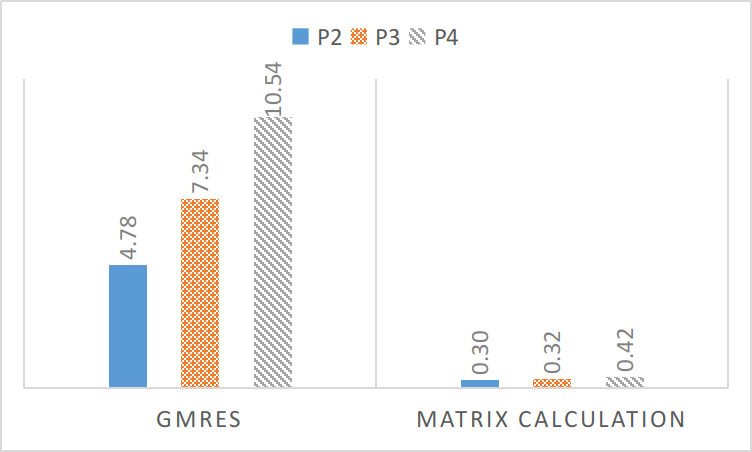}
							\par\end{centering}
						\begin{centering}
							\par\end{centering}
					}
					
					\subfloat[Operations in GMRES \label{fig:PerfAnaly-3D-GMRES}]{\begin{centering}
							\includegraphics[width=0.5\textwidth]{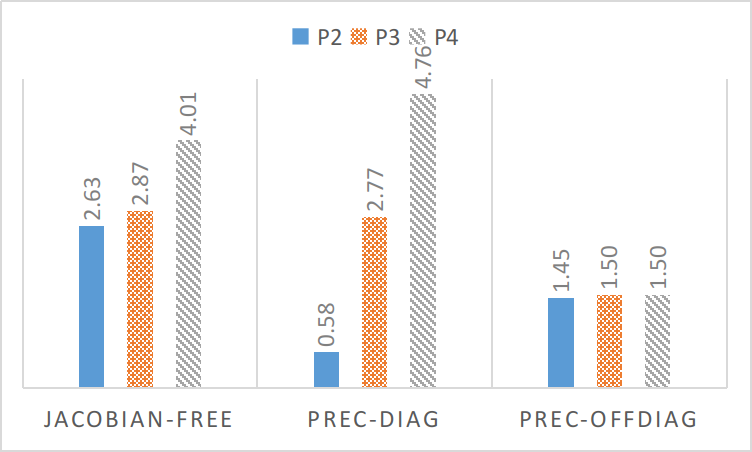}
							\par\end{centering}
						\begin{centering}
							\par\end{centering}
					}\subfloat[Operations of matrix calculation  of preconditioner\label{fig:PerfAnaly-3D-Prec}]{\begin{centering}
							\includegraphics[width=0.5\textwidth]{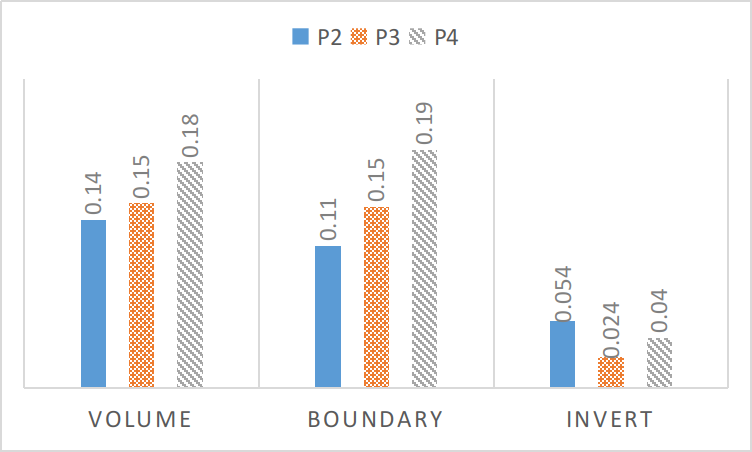}
							\par\end{centering}
						\begin{centering}
							\par\end{centering}
					}
					\par\end{centering}
				\begin{centering}
					\caption{Performance analysis of the implicit solver in 2D \label{fig:PerfAnaly-3D}}
					\par\end{centering}
			\end{figure}

		\subsubsection{2D case} \label{performance-analy-2D}
			The 2D cylinder test cases of Section \ref{2D_cylinder} are adopted for the performance analysis of the solver in 2D. The simulations are run at CFL=20 for 200 steps with preconditioning matrices updated every 10 time steps. The other parameters are the same as those of the 3D tests.	 Fig. \ref{fig:PerfAnaly-3D} shows the CPU time of different parts of the implicit solver, which are normalized in the same way. Very similar trends as in the 3D case are observed. The $\hat{\mathbf{D}}^{-1}$ vector inner product still grows much faster than the other parts, which means matrix-based preconditioners may also lose efficiency because of the large computational   cost. However, the growth of CPU time in preconditioning matrices calculation is much slower than that in 3D case. The CPU time in inverting the $\hat{\mathbf{D}}$ matrix is not increasing monotonously as that of the 3D case. A possible reason is that the functions in intel MKL library are sensitive to the rank of the matrix when the matrix size is small, as discussed in reference \citep{hillewaert_development_2013}.
 
	\subsection{Memory consumption analysis}\label{sec:Memory-analysis}
	
		Tab.~\ref{tab:BRJ-Memory-consumption} compares the memory requirements of the preconditioning matrices and Jacobian matrix $\partial\mathbf{N}/\partial\mathbf{u}$ using the 3D cylinder case of Section \ref{performance-analy-3D}. The memory consumption of the Jacobian-Sparse is that of the Jacobian matrix considering the sparsity patterns of the off diagonal blocks of $\partial\mathbf{N}/\partial\mathbf{u}$, which is calculated based on the estimation in reference \citep{peraire_compact_2008}. The memory consumption of the Jacobian-Dense is used by storing the diagonal and off diagonal blocks as dense matrices. Note that the ILU preconditioner with zero fill-in should have the same memory consumption as the Jacobian-Dense since the matrices will generally lose sparsity after the factorization. The memory consumption of the Jacobian matrix becomes large in 3D. Only approximately 45 elements
		can be afforded per Gbyte of memory for simulations with $P=4$ and
		hexahedral meshes if the Jacobian-Dense matrix is stored or ILU preconditioner is used. Moreover, frequently loading these large matrices into the CPU cache will seriously influence
		the computational speed since most modern high-performance computers
		(HPCs) are memory bandwidth limited \citep{williams_roofline:_2009}. The memory consumption can be further reduced if using other schemes like the CDG \citep{peraire_compact_2008} by exploring the sparsity patterns, but it is still of the same order of magnitude as the Jacobian-Sparse. 

		Memory in operations BRJ-Diag and BRJ-Offdiag is required for storing $\hat{\mathbf{D}}^{-1}$ and $\partial\hat{\boldsymbol{\mathbf{H}}}^{n}/\partial\hat{\mathbf{Q}}^{\pm}$, both of which are stored in single precision to reduce memory and CPU costs. The storage of $\partial\hat{\boldsymbol{\mathbf{H}}}^{n}/\partial\hat{\mathbf{Q}}^{\pm}$ is small compared with that of $\hat{\mathbf{D}}^{-1}$ especially for high order approximations. The total storage of BRJ is much smaller than that of the Jacobian matrix or the ILU preconditioner.

		\begin{table}[H]
			\begin{centering}
			\begin{tabular}{|c||r|r|r|r|r|}
			\hline 
			 & \textbf{\scriptsize{}BRJ-Diag} & \textbf{\scriptsize{}BRJ-Offdiag} & \textbf{\scriptsize{}BRJ-Total} & \textbf{\scriptsize{}Jacobian-Dense} & \textbf{\scriptsize{}Jacobian-Sparse}\tabularnewline
			\hline 
			\hline 
			\textbf{\scriptsize{}$P=2$} & {\scriptsize{}72 900} & {\scriptsize{}30 000} & {\scriptsize{}102 900} & {\scriptsize{}1 020 600} & {\scriptsize{}712 800}\tabularnewline
			\hline 
			\textbf{\scriptsize{}$P=3$} & {\scriptsize{}409 600} & {\scriptsize{}43 200} & {\scriptsize{}452 800} & {\scriptsize{}5 734 400} & {\scriptsize{}3 328 000}\tabularnewline
			\hline 
			\textbf{\scriptsize{}$P=4$} & {\scriptsize{}1 562 500} & {\scriptsize{}76 800} & {\scriptsize{}1 639 300} & {\scriptsize{}21 875 000} & {\scriptsize{}11 000 000}\tabularnewline
			\hline 
			\end{tabular}
			\par\end{centering}
			\caption{Memory consumption of BRJ in bytes for each hexahedral mesh\label{tab:BRJ-Memory-consumption}}
		\end{table}

\section{Verification and applications} \label{sec:V-and-V}

	In the following, the implementations of spatial discretization
	methods are verified using accuracy tests described in Section \ref{sec:accuracy-test}. Section
	\ref{2D_cylinder} discusses the temporal accuracy and efficiency of the solver in
	unsteady laminar flows over a circular cylinder. The temporal accuracy
	in turbulent simulations is studied in Section \ref{TGV} using a Taylor-Green
	vortex case. Section \ref{3D_cylinder} presents and discusses turbulent flow over a
	circular cylinder at $\text{Re}=3900$ obtained with the implicit
	solver. Finally, a shock wave boundary layer interaction is studied
	to demonstrate the shock capturing ability of the compressible flow solver.

	In this section, we use	two different definitions of
	Courant-Friedrichs-Lewy (CFL) number which are important for the choices of
	time steps, are presented. The "standard" CFL number is defined as
	\begin{equation}
		\text{CFL}=\frac{c_{\lambda}\triangle t\left(c+\sqrt{u_{k}u_{k}}\right)P^{2}}{d_{RK}\triangle x},\label{eq:CFL}
	\end{equation}
	where $c_{\lambda}=0.2$, $d_{RK}=2$ and $c$ is the speed of sound.
	The time step of the explicit solver will be chosen based on
	Eq. \eqref{eq:CFL}, which ensures the maximum stable $\text{CFL}$ is
	almost constant for different values of the polynomial order $P$ \citep{karniadakis_spectral/hp_2013}.
	Another CFL number, the convective CFL number, is defined as
	\begin{equation}
		\text{CFL}_{c}=\frac{\triangle t\sqrt{u_{k}u_{k}}}{\triangle x/P}.\label{eq:CFLc}
	\end{equation}
	A value $\text{CFL}_{c}=1$ means the flow field can convect a distance
	of $\triangle x/P$ in one time step, where the distance $\triangle x/P$ in
	Eq. \eqref{eq:CFLc} is an estimate of the average length  between two
	DoFs in the element. The time step of the implicit solver maintains a small $\text{CFL}_{c}$, which ensures the temporal
	error is similar to or smaller than the spatial error.

	\subsection{Accuracy test} \label{sec:accuracy-test}

		Testing the convergence order of accuracy (OoA) of the solver is a
		very effective way of code verification \citep{roy_review_2005}.
		Here two different test cases are used to verify two different aspects
		of the solver. An isentropic vortex convection problem is used to verify
		the advection scheme and the treatment of the diffusion terms is verified
		by means of a manufactured compressible Poiseuille
		flow.

		\begin{figure}[H]
			\begin{centering}
				\includegraphics[width=0.75\textwidth]{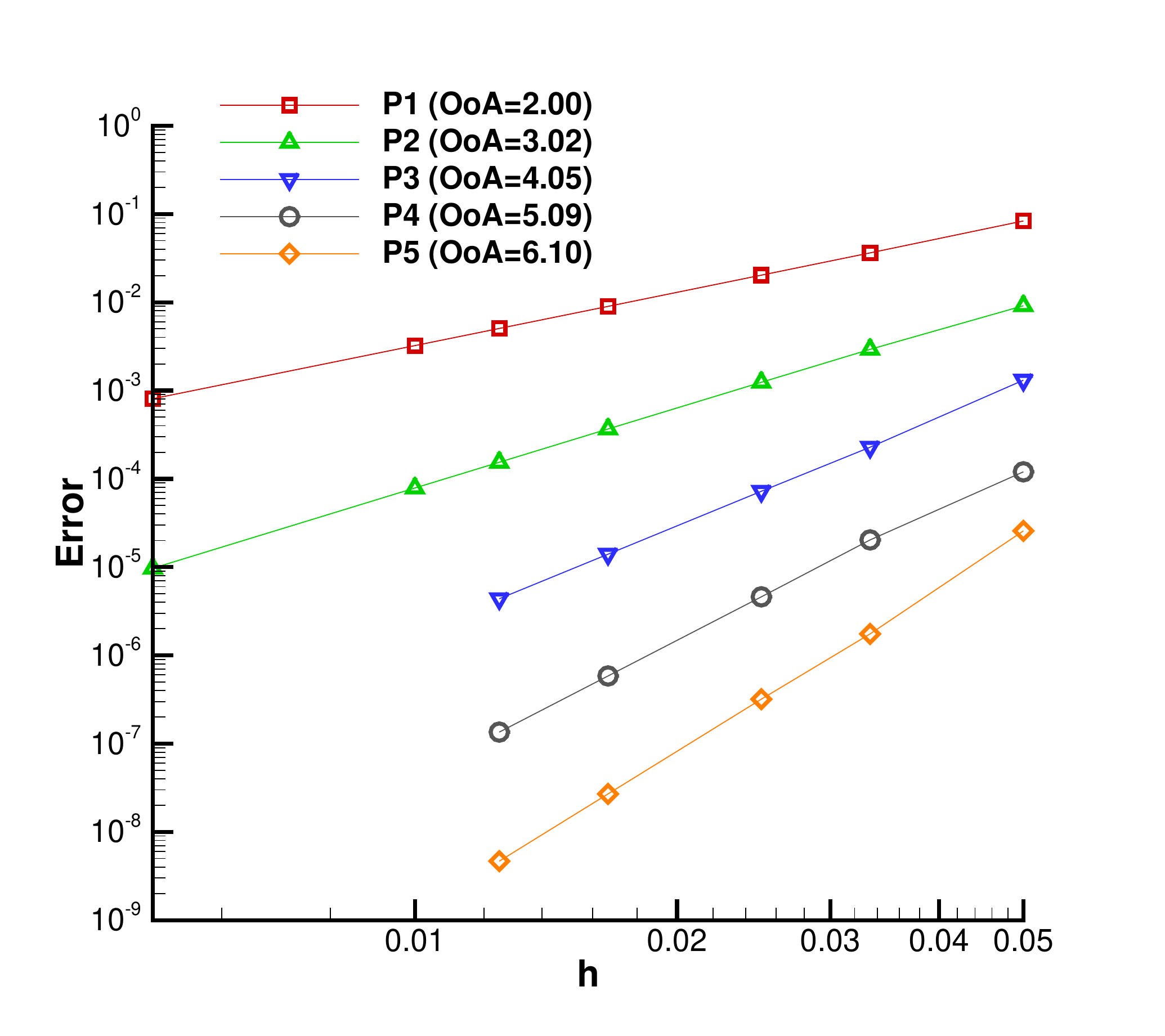}
				\par\end{centering}
			\centering{}\caption{Accuracy test of the advection scheme \label{fig:Accuracy-Isentropic-Vortex}}
		\end{figure}

		\paragraph{Isentropic vortex convection}

		In this case an inviscid isentropic vortex is convected down stream.
		In computational domain $[0,10]^2$, the exact solution at time $t$ is
		\begin{equation}
			\begin{aligned}\rho= & \left(1-\frac{\varphi^{2}\left(\gamma-1\right)}{16\gamma\pi^{2}}e^{2\left(1-r^{2}\right)}\right)^{\frac{1}{\gamma-1}} \\
				u=    & u_{0}+\frac{\varphi\left(y-y_{0}\right)}{2\pi}e^{\left(1-r^{2}\right)}                                                \\
				v=    & v_{0}+\frac{\varphi\left(x-x_{0}\right)}{2\pi}e^{\left(1-r^{2}\right)}                                                \\
				p=    & \rho^{\gamma}
			\end{aligned}
			,
		\end{equation}
		where $\left(u_{0},v_{0}\right)=\left(1.0,0.5\right)$ is the mean
		convection velocity field,
		the coordinates of the vortex center at time $t$ are given by $\left(x_{0}+u_{0}t,y_{0}+v_{0}t\right)$, $r$ is the distance from
		the vortex center, and $\varphi=5.0$ is a parameter that controls the strength of the vortex.
		Mesh convergence is analyzed through the use of progressively
		refined quadrilateral meshes. Periodic boundary conditions
		are applied to the boundaries in both directions. A fourth-order explicit Runge-Kutta with CFL=0.01 is adopted to guarantee the errors are dominated by the spatial
		discretization. The error distributions are depicted in Fig. \ref{fig:Accuracy-Isentropic-Vortex}
		with the order of accuracy (OoA) between the two finest mesh levels given in the legend. All simulations with polynomial orders $P$ from 1 to 5 reach the designed
		order of accuracy.

		\paragraph{Manufactured compressible Poiseuille flow}

\begin{figure}[H]
			\begin{centering}
				\includegraphics[width=0.75\textwidth]{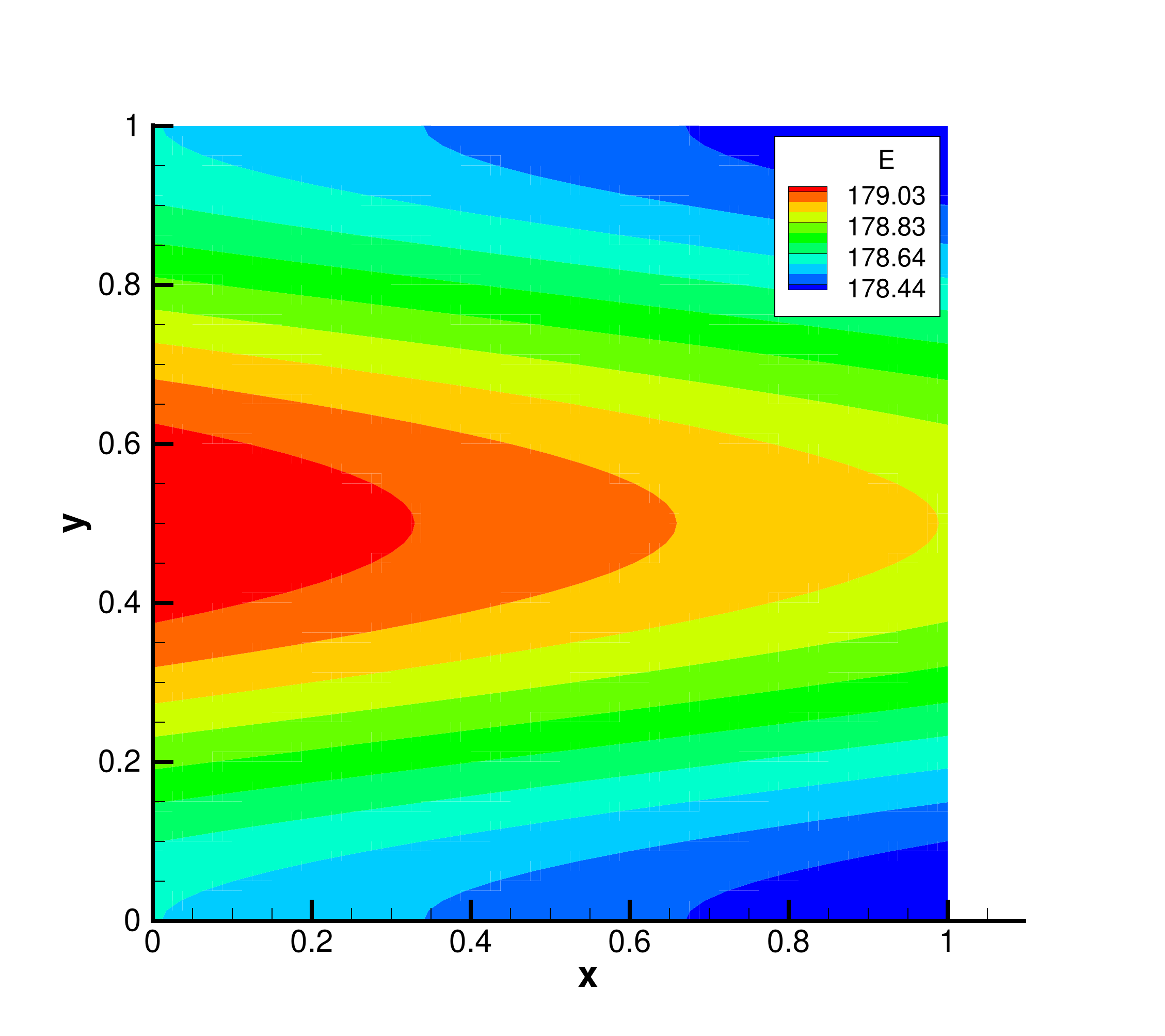}
				\par\end{centering}
			\centering{}\caption{Total energy ($E$) distribution of the manufactured compressible
				Poiseuille flow\label{fig:Total-energy-distribution-Poiseuille's}}
		\end{figure}

\begin{figure}[H]
			\begin{centering}
				\includegraphics[width=0.5\textwidth]{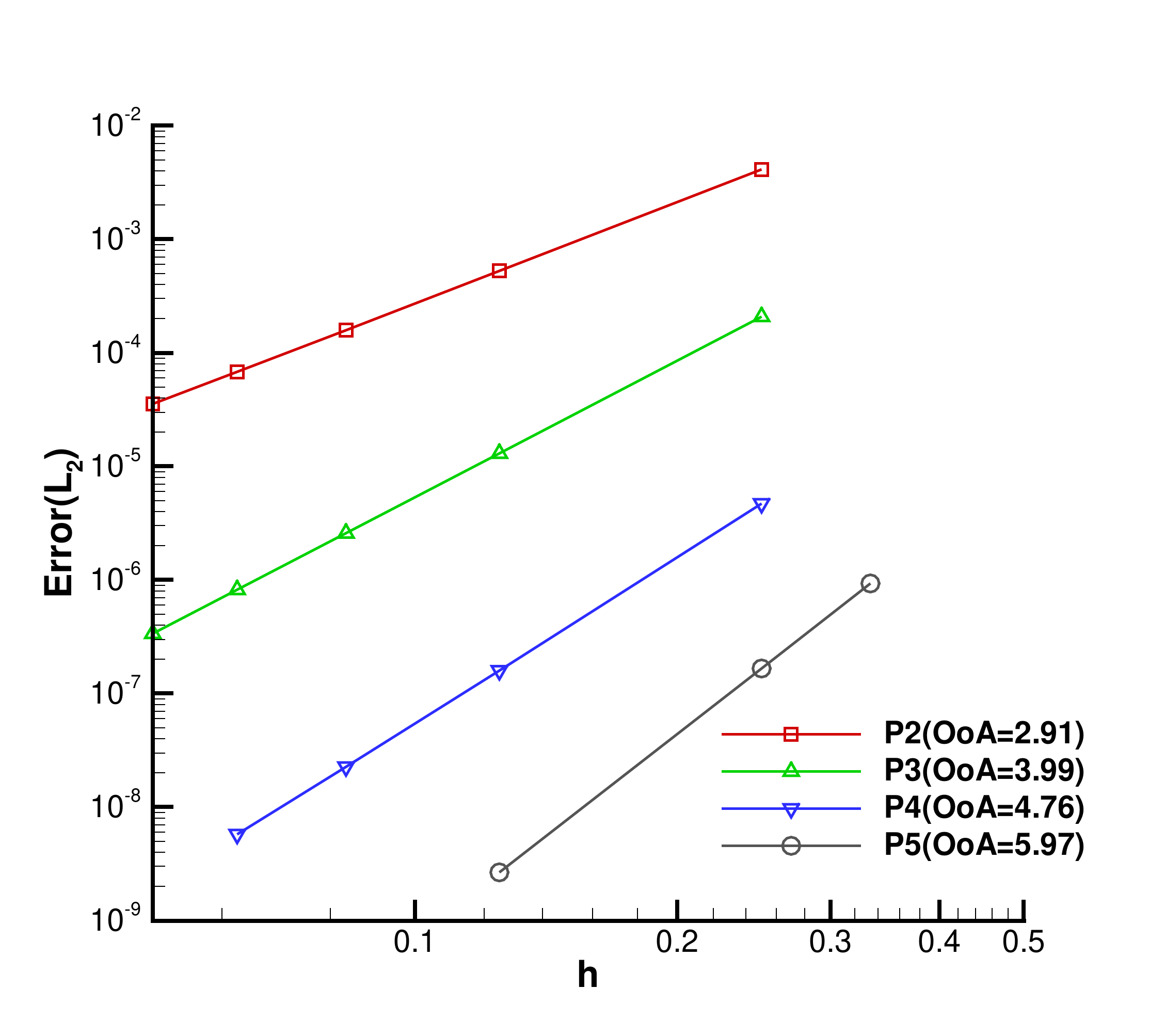}\includegraphics[width=0.5\textwidth]{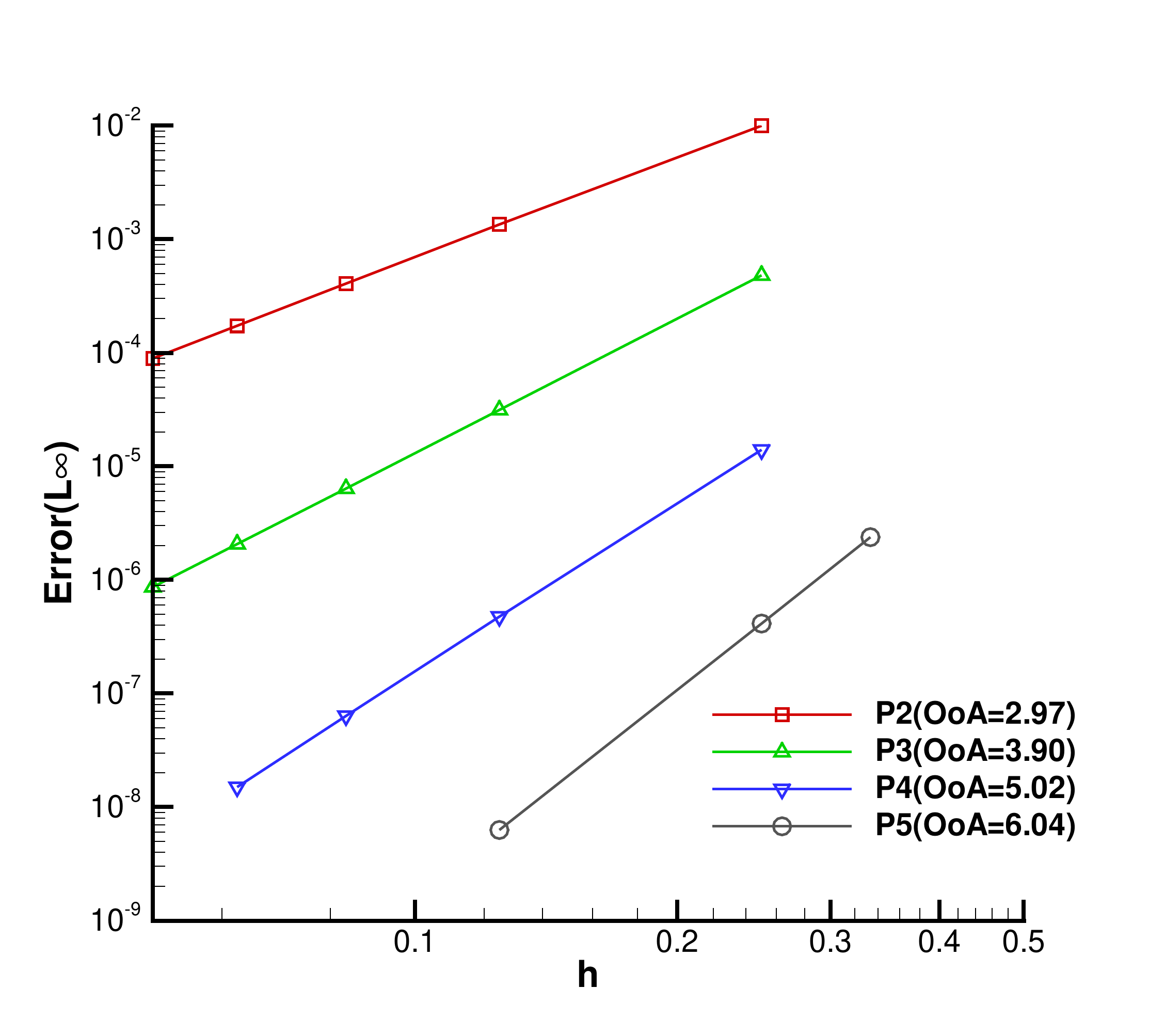}
				\par\end{centering}
			\centering{}\caption{Convergence order of the IP methods in compressible Poiseuille
				flow\label{fig:Convergence-order-of-Poiseuille's}}
		\end{figure}

		The method of manufactured solution (MMS) permit us to design specific
		test cases with analytical solutions for the study of different aspects
		of the solvers \citep{roy_review_2005}. Using the MMS, a test case
		similar to Poiseuille flow \citep{oliver_multigrid_2004} is designed
		to verify the DG methods for the diffusion terms.  This problem
		is suitable for the verification of the IP method since it is dominated
		by the diffusion term. By design, the analytical solution
		is

		\begin{equation}
			\begin{aligned}\rho & =1                                                                                           \\
				u    & =-\frac{1}{2\mu}\frac{dp}{dx}\left[y\left(L-y\right)+\theta y^{2}\left(L-y\right)^{2}\right] \\
				v    & =0                                                                                           \\
				p    & =\frac{1}{\gamma\text{Ma}^{2}}+\frac{dp}{dx}x
			\end{aligned}
			,\label{eq:MMS-SOLUTION}
		\end{equation}		
		where $L=1$ is the height of the computational domain in the $y$
		direction. The corresponding forcing terms are given in Eqs. \eqref{eq:forcing-MMS}
		in the Appendix and correspond to a free-stream flow with $\text{Ma}=0.1$ and $\text{Re}=100$ based on the maximum velocity, which is normalized to 1. The corresponding pressure gradient is $dp/dx=-8\mu/L^2$. Compared
		to incompressible Poiseuille flow, an extra term $\theta y^{2}\left(L-y\right)^{2}$
		with $\theta=0.01$ is added so that the analytical solution of the energy $E$, shown in Fig.~\ref{fig:Total-energy-distribution-Poiseuille's}, 
		has a polynomial order of $8$.
		This ensures the polynomial order
		of the solution is higher than that of the base functions and thus avoids
		the high-order coefficients becoming excessively small. A series of
		quadrilateral meshes with $3^{2}$, $4^{2}$, $8^{2}$, $12^{2}$, $16^{2}$
		and $20^{2}$ elements are used with Dirichlet boundary conditions
		based on the analytical solution.
		The $L_{2}$ and $L_{\infty}$ norm error distributions
		are shown in Fig.~\ref{fig:Convergence-order-of-Poiseuille's}. The results show that the designed order of accuracy
		is achieved for the IP method from $P=2$ to $P=5$. Simulations with
		triangular meshes can also achieve the designed order of accuracy but
		are not shown here.

		\begin{figure}[H]
			\begin{centering}
				\includegraphics[width=0.5\textwidth]{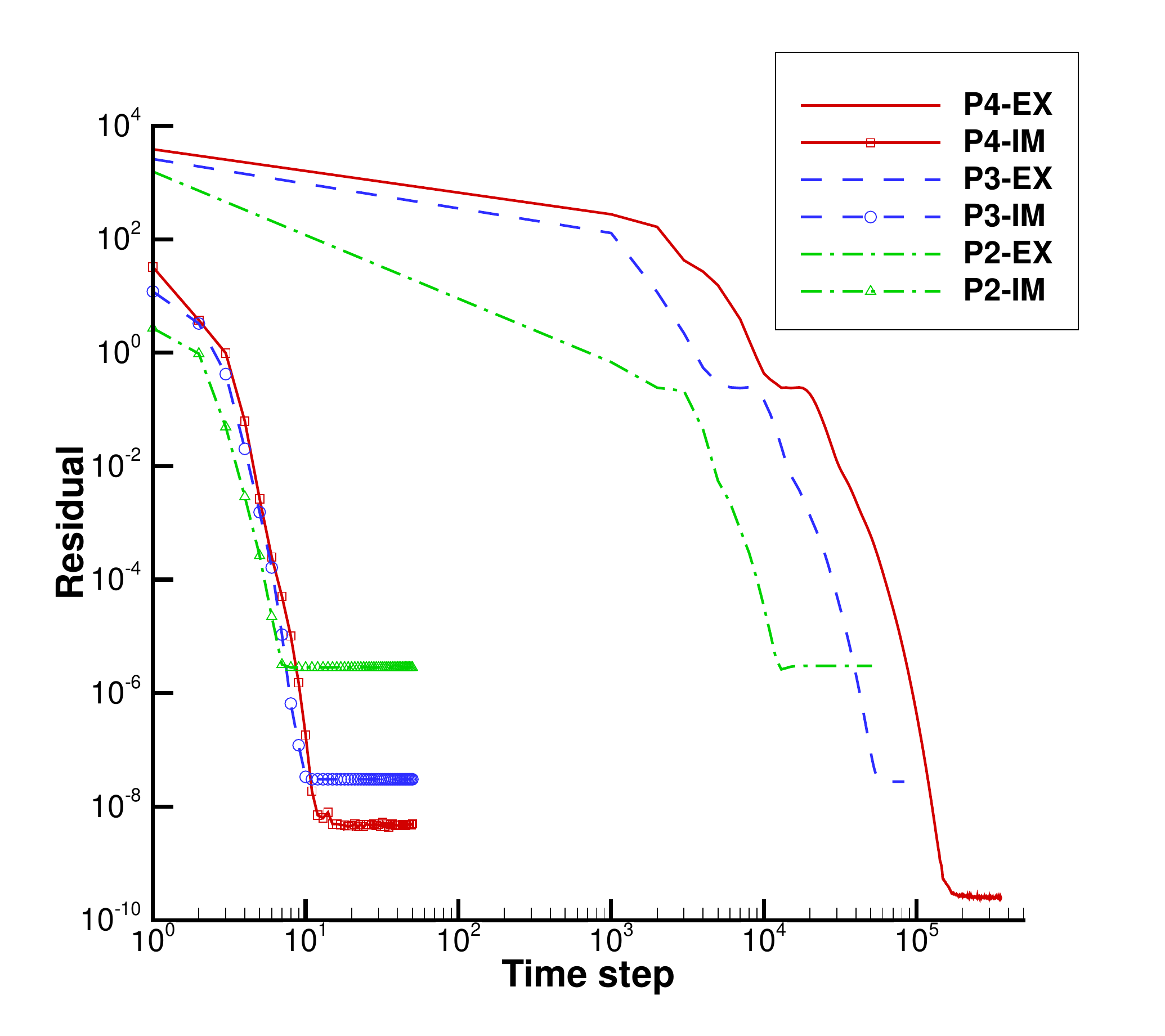}\includegraphics[width=0.5\textwidth]{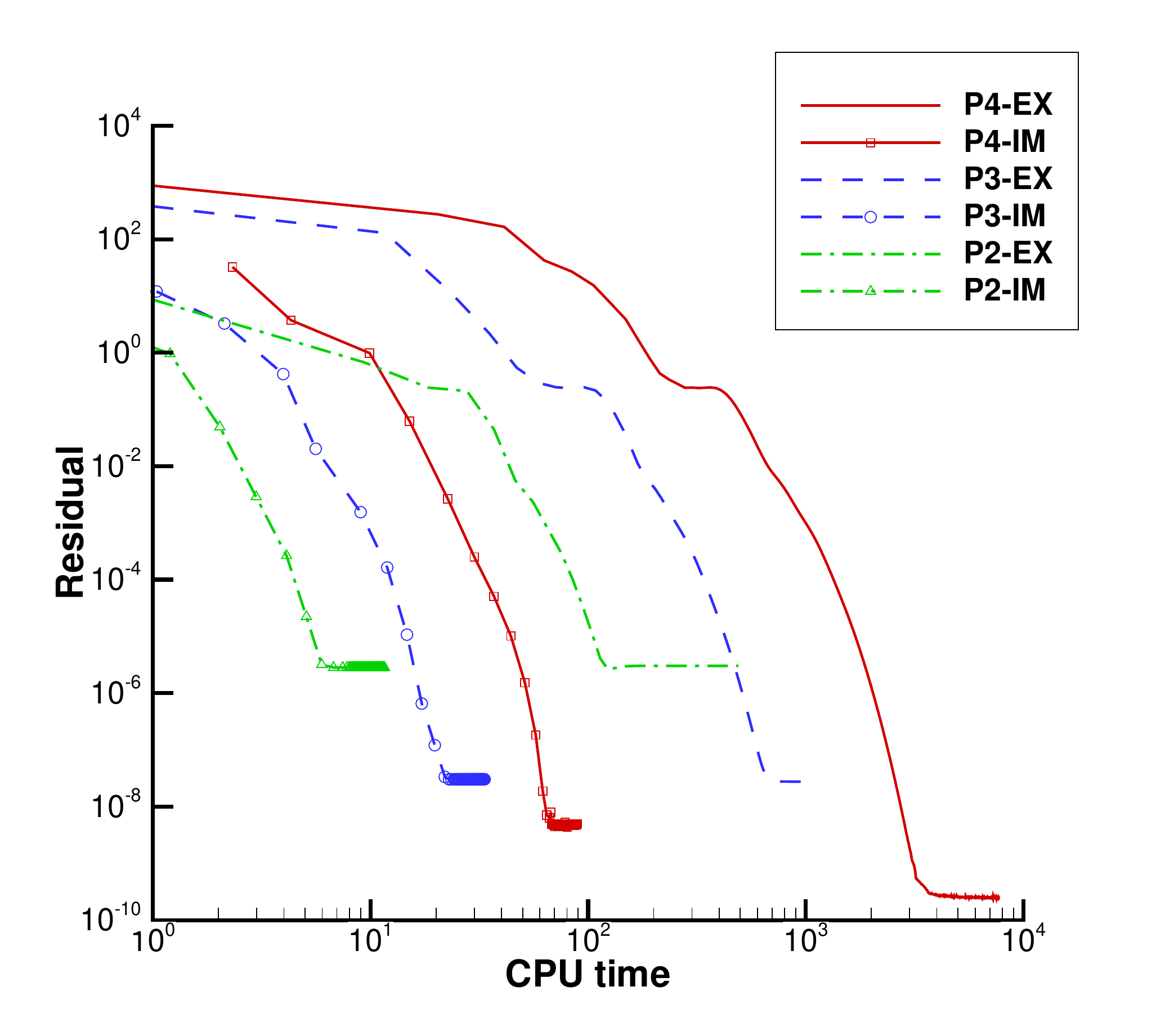}
				\par\end{centering}
			\begin{centering}
				\caption{Residual distribution of implicit and explicit simulations in compressible
					Poiseuille flow\label{fig:Residual-distribution-of}}
				\par\end{centering}
		\end{figure}

		The efficiency of the two solvers is also compared in this steady problem.
		The initial condition is similar to that used in Eqs. \eqref{eq:MMS-SOLUTION}
		but with uniform velocity $u=-(dp/dx)L^{2}/(8\mu)$. The
		time integration used by the implicit solver is SDIRK2 whilst the explicit solver uses ERK2. The
		CFL number of the implicit solver grows from 500 to 5000 within the first 10
		steps. Here the preconditioning matrices are updated in every other time step.
		The explicit solver uses its maximum stable CFL numbers of $0.08$, $0.06$ and $0.06$ 
		for $P=2$, $P=3$ and $P=4$, respectively. Fig.~\ref{fig:Residual-distribution-of}
		shows the convergence history of the residual norm
		$\left\Vert \mathbf{M}^{-1}\mathbf{\mathcal{\mathcal{L}}}\right\Vert _{2}$.
		All the implicit simulations reach steady state
		within 20 steps, and the explicit solver takes about 13000, 60000 and 170000 steps
		for $P=2$, $P=3$ and $P=4$ to converge, respectively. Fig. \ref{fig:Residual-distribution-of}
		shows that even with no grid stretching the implicit solver is about
		$18.2$, $33.1$ and $53.4$ times faster than the explicit solver for $P=2$,  $P=3$
		and $P=4$, respectively.

		Following references \citep{oliver_multigrid_2004,roy_review_2005}, Dirichlet boundary conditions based on the analytical solution are applied weakly in the DGM in this subsonic problem. The difference between analytical solutions and numerical solutions will lead to large errors near boundaries, which prevents the simulations from converging to machine epsilon levels. However, the errors caused by boundary treatment does not invalidate the OoA tests since the difference between the numerical and analytical solutions on the boundaries tends
		to zero as the resolution increases.

	\subsection{Laminar flow over a circular cylinder} \label{2D_cylinder}

\begin{figure}[H]
			\begin{centering}
				\includegraphics[width=0.5\textwidth]{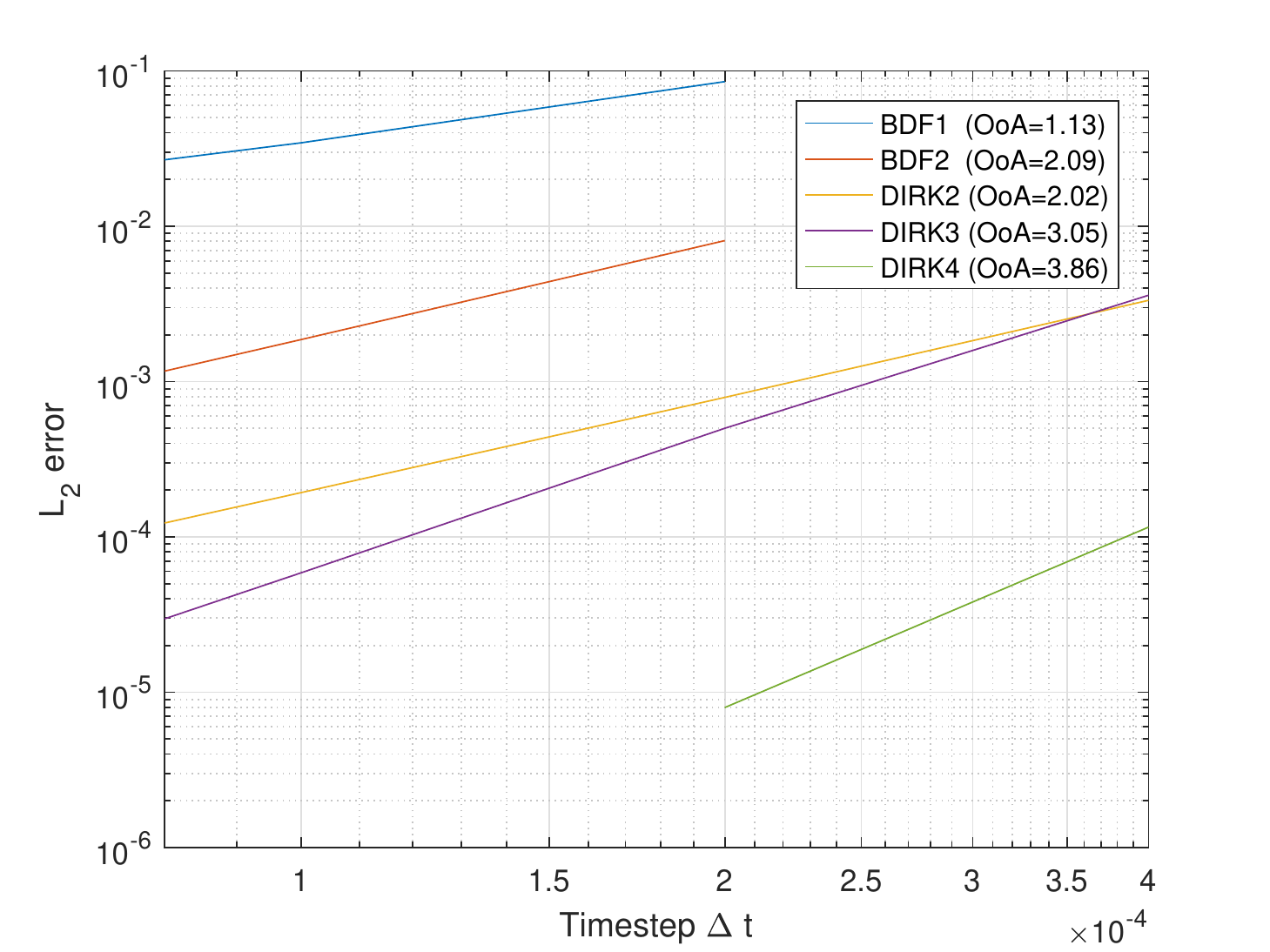}\includegraphics[width=0.5\textwidth]{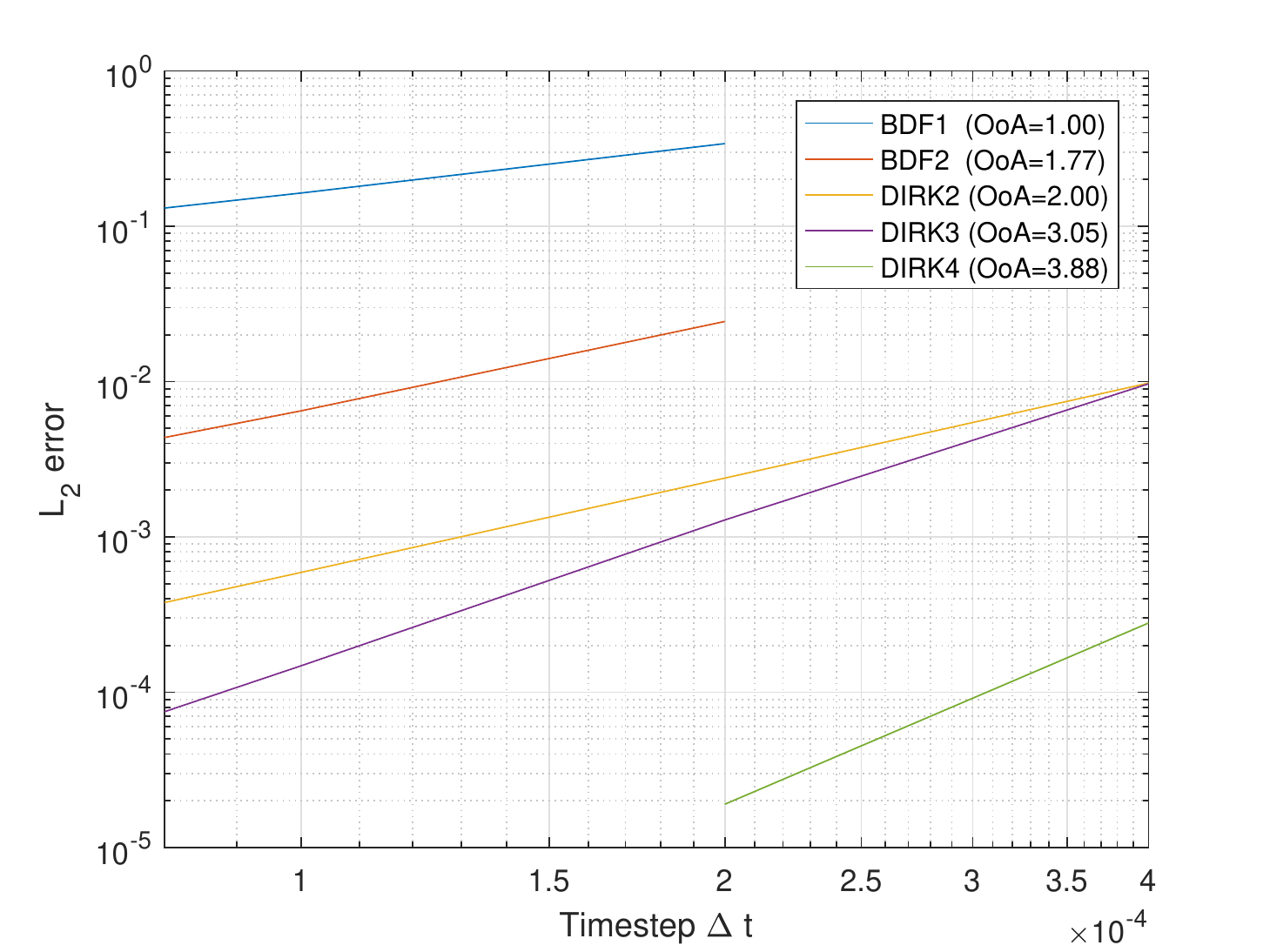}
				\par\end{centering}
			\centering{}\caption{Convergence rate of different implicit time integration schemes\label{fig:Convergence-rate-of-Vortex-shedding}}
		\end{figure}

		The temporal accuracy and efficiency of the implicit solver is tested next
		using unsteady simulations of flow over a circular cylinder. The parameters and meshes employed are
		similar to those presented in reference \citep{bijl_implicit_2002}. The flow conditions correspond to
		$\text{Ma}=0.3$, $\text{Re}=\rho uD/\mu=1200$ and we use
		$64\times60$ quadrilateral meshes. Starting from the same flow field,
		different time integration schemes and time steps are used to get
		the solutions after approximately 1.7 vortex shedding periods. The
		integrated lift and drag forces are the chosen metrics for the accuracy
		test with reference solutions obtained by the SDIRK4 in Tab. 16 of
		reference \citep{kennedy_diagonally_2016} with a very small time
		step $\Delta t=1\times10^{-5}$. A smaller Newton iteration tolerance $\alpha=10^{-6}$ is used for SDIRK3 and SDIRK4. The convergence rates are presented in
		Fig.~\ref{fig:Convergence-rate-of-Vortex-shedding}, which shows that the
		implicit solver achieves the desired order of accuracy.

		The efficiency of the implicit solver is compared with that of the
		explicit solver in this test case. Starting from the same flow field,
		the solution is obtained after around $30$ vortex shedding periods
		using both explicit and implicit solver. The details of the simulations
		are summarized in Tab.~\ref{tab:Efficiency-comparison-between}. The
		implicit solver can run with a time step 330 times larger than that
		of the explicit solver. The corresponding $\text{CFL}_{c}$ is about
		$2.5$ for this time step. The implicit solver is around 20.3 times
		quicker than the explicit solver. The implicit solver is also ran
		with a larger time step ($\text{CFL}_{c}\simeq5$), which achieves
		a speed-up of 27.9. By observing the evolution history of the
		flow field, it can be observed that the temporal errors mainly lead to a dispersion
		in the vortex shedding process. Thus the Strouhal number ($\text{Sr}=\frac{Df}{u}$ with $f$  the vortex shedding frequency) is
		a good metric to compare the temporal accuracy. Tab.~\ref{tab:Efficiency-comparison-between}
		shows that all the simulations give almost the same prediction of
		Strouhal number (0.2419). The results show that the implicit solver
		can largely improve efficiency while keeping good temporal accuracy.

		\begin{table}[H]
			\begin{centering}
				\begin{tabular}{|c|c|c|c|}
					\hline
					\textbf{\scriptsize{}Method}     & \textbf{\scriptsize{}ERK2}                      & \multicolumn{2}{c|}{\textbf{\scriptsize{}SDIRK2}}\tabularnewline
					\hline
					\hline
					\textbf{\scriptsize{}$\Delta t$} & {\scriptsize{}$4.5\times10^{-5}$}              & {\scriptsize{}$1.5\times10^{-2}$}                               & {\scriptsize{}$2.0\times10^{-2}$}\tabularnewline
					\hline
					{\scriptsize{}$\text{CFL}$}      & {\scriptsize{}0.06}                            & {\scriptsize{}20}                                               & {\scriptsize{}40}\tabularnewline
					\hline
					{\scriptsize{}$\text{CFL}_{c}$}  & {\scriptsize{}$7.5\times10^{-3}$}              & {\scriptsize{}2.5}                                              & {\scriptsize{}5.0}\tabularnewline
					\hline
					{\scriptsize{}Strouhal number}   & {\scriptsize{}0.2419}                          & {\scriptsize{}0.2419}                                           & {\scriptsize{}0.2419}\tabularnewline
					\hline
					{\scriptsize{}Newton iterations} & \multirow{2}{*}{{\scriptsize{}\textbackslash}} & \multirow{2}{*}{{\scriptsize{}2.1}}                               & \multirow{2}{*}{{\scriptsize{}2.5}}\tabularnewline
					{\scriptsize{}per RK stage}      &                                                &                                                                 & \tabularnewline
					\hline
					{\scriptsize{}GMRES iterations}  & \multirow{2}{*}{{\scriptsize{}\textbackslash}} & \multirow{2}{*}{{\scriptsize{}3.5}}                              & \multirow{2}{*}{{\scriptsize{}5.0}}\tabularnewline
					{\scriptsize{}per RK stage}      &                                                &                                                                 & \tabularnewline
					\hline
					{\scriptsize{}Speed-up}          & {\scriptsize{}1.0}                             & {\scriptsize{}20.3}                                             & {\scriptsize{}27.9}\tabularnewline
					\hline
				\end{tabular}
				\par\end{centering}
			\caption{Efficiency comparison between explicit and implicit time integration
				schemes\label{tab:Efficiency-comparison-between}}
		\end{table}

		\begin{table}[H]
			\begin{centering}
			\begin{tabular}{|c||c|c||c|c|}
			\hline 
			\textbf{\scriptsize{}Polynomial order} & \multicolumn{2}{c||}{$P=2$} & \multicolumn{2}{c|}{$P=3$}\tabularnewline
			\hline 
			\textbf{\scriptsize{}Method} & \textbf{\scriptsize{}ERK2} & \textbf{\scriptsize{}SDIRK2} & \textbf{\scriptsize{}ERK2} & \textbf{\scriptsize{}SDIRK2}\tabularnewline
			\hline 
			\hline 
			\textbf{\scriptsize{}$\Delta t$} & {\scriptsize{}$1.1\times10^{-3}$} & {\scriptsize{}$0.1$} & {\scriptsize{}$6.1\times10^{-4}$} & {\scriptsize{}$0.046$}\tabularnewline
			\hline 
			{\scriptsize{}$\text{CFL}$} & {\scriptsize{}0.08} & {\scriptsize{}7.5} & {\scriptsize{}0.10} & {\scriptsize{}7.5}\tabularnewline
			\hline 
			{\scriptsize{}$\text{CFL}_{c}$} & {\scriptsize{}$2.6\times10^{-2}$} & {\scriptsize{}2.5} & {\scriptsize{}$1.8\times10^{-2}$} & {\scriptsize{}1.4 }\tabularnewline
			\hline 
			\textbf{\scriptsize{}Newton iterations} & \multirow{2}{*}{{\scriptsize{}\textbackslash{}}} & \multirow{2}{*}{{\scriptsize{}4}} & \multirow{2}{*}{{\scriptsize{}\textbackslash{}}} & \multirow{2}{*}{{\scriptsize{}5}}\tabularnewline
			\textbf{\scriptsize{}per RK stage} &  &  &  & \tabularnewline
			\hline 
			\textbf{\scriptsize{}GMRES iterations} & \multirow{2}{*}{{\scriptsize{}\textbackslash{}}} & \multirow{2}{*}{{\scriptsize{}3.7}} & \multirow{2}{*}{{\scriptsize{}\textbackslash{}}} & \multirow{2}{*}{{\scriptsize{}4.6}}\tabularnewline
			\textbf{\scriptsize{}per Newton iteration} &  &  &  & \tabularnewline
			\hline 
			\textbf{\scriptsize{}Speed-up} & {\scriptsize{}1.0} & {\scriptsize{}2.76} & {\scriptsize{}1.0} & {\scriptsize{}1.2}\tabularnewline
			\hline 
			\end{tabular}
			\par\end{centering}
			\caption{Efficiency comparison between explicit and implicit time integration
			schemes of TGV \label{tab:Efficiency-comparison-between-2-1}}
		\end{table}

		\begin{figure}[H]
			\begin{centering}
				\subfloat[Different resolutions (variable $P$) \label{fig:Different-resolutions}]{\begin{centering}
						\includegraphics[width=0.5\textwidth]{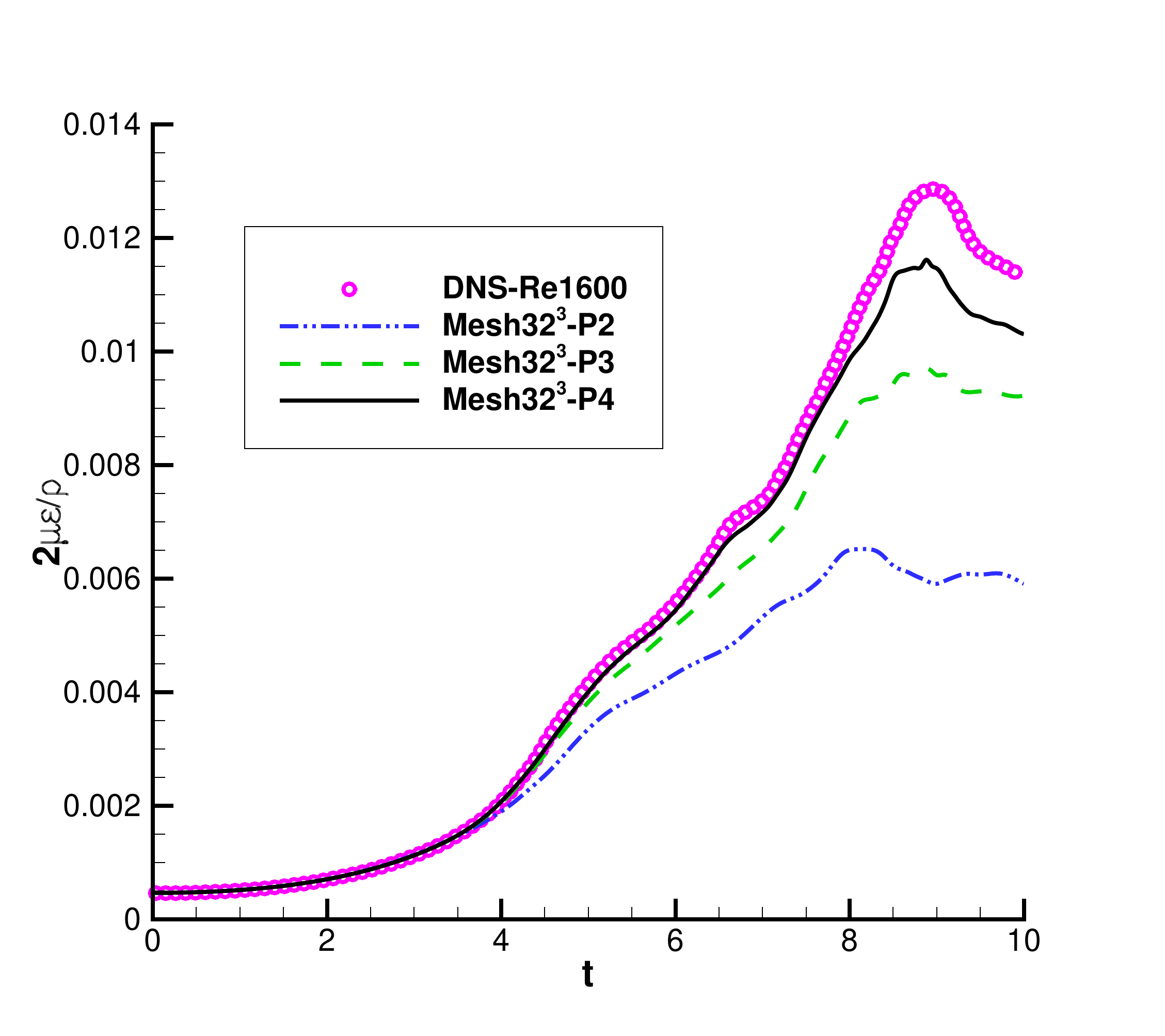}
						\par\end{centering}
					\begin{centering}
						\par\end{centering}
				}\subfloat[Different temporal methods\label{fig:Different-temporal-methods}]{\begin{centering}
						\includegraphics[width=0.5\textwidth]{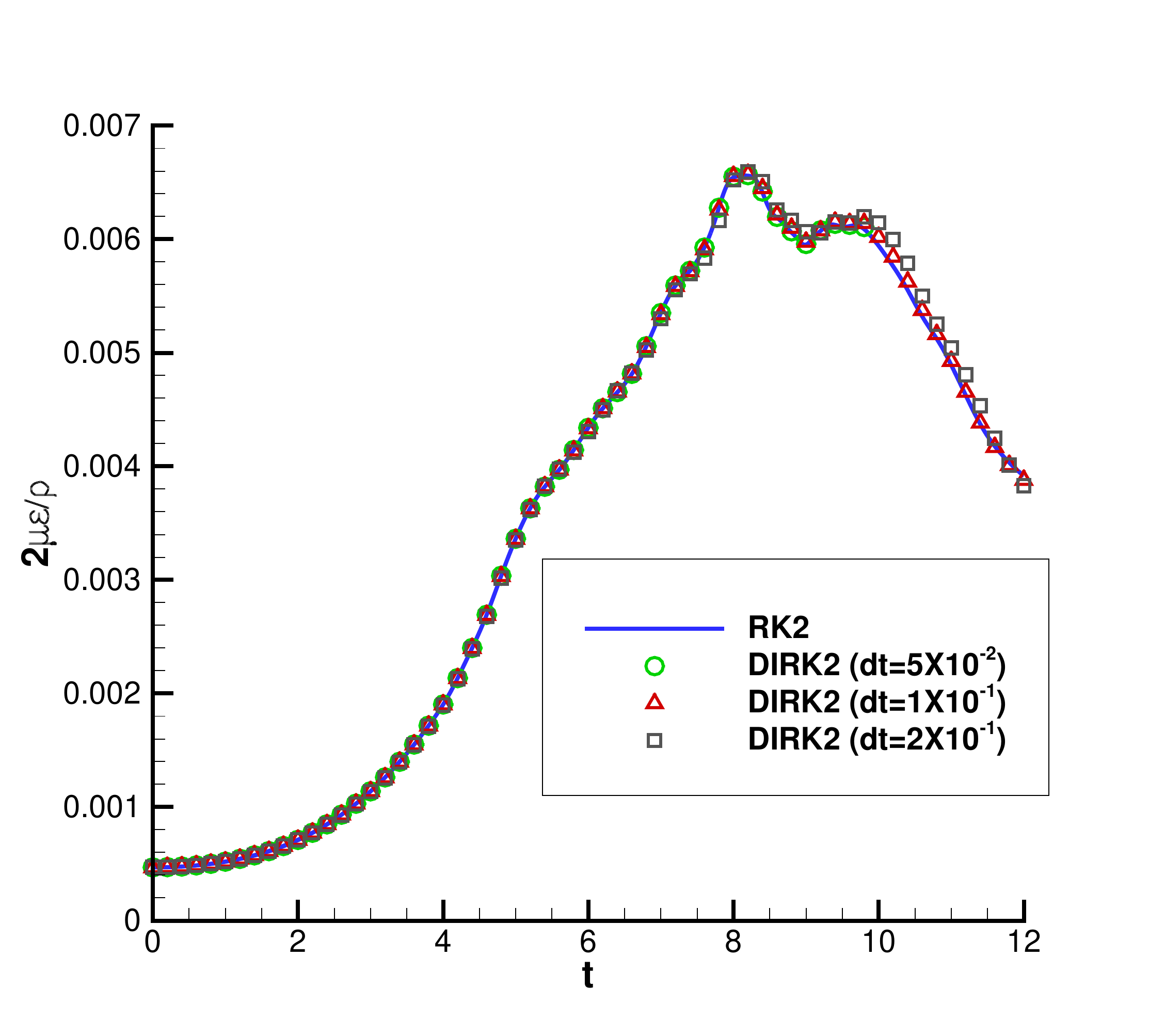}
						\par\end{centering}
					\begin{centering}
						\par\end{centering}
				}
				\par\end{centering}
			\begin{centering}
				\caption{Evolution of the dissipation rate in Taylor-Green vortex \label{fig:Evolution-of-the}}
				\par\end{centering}
		\end{figure}

	\subsection{Taylor-Green vortex} \label{TGV}

\begin{figure}[H]
			\begin{centering}
				\includegraphics[width=0.7\textwidth]{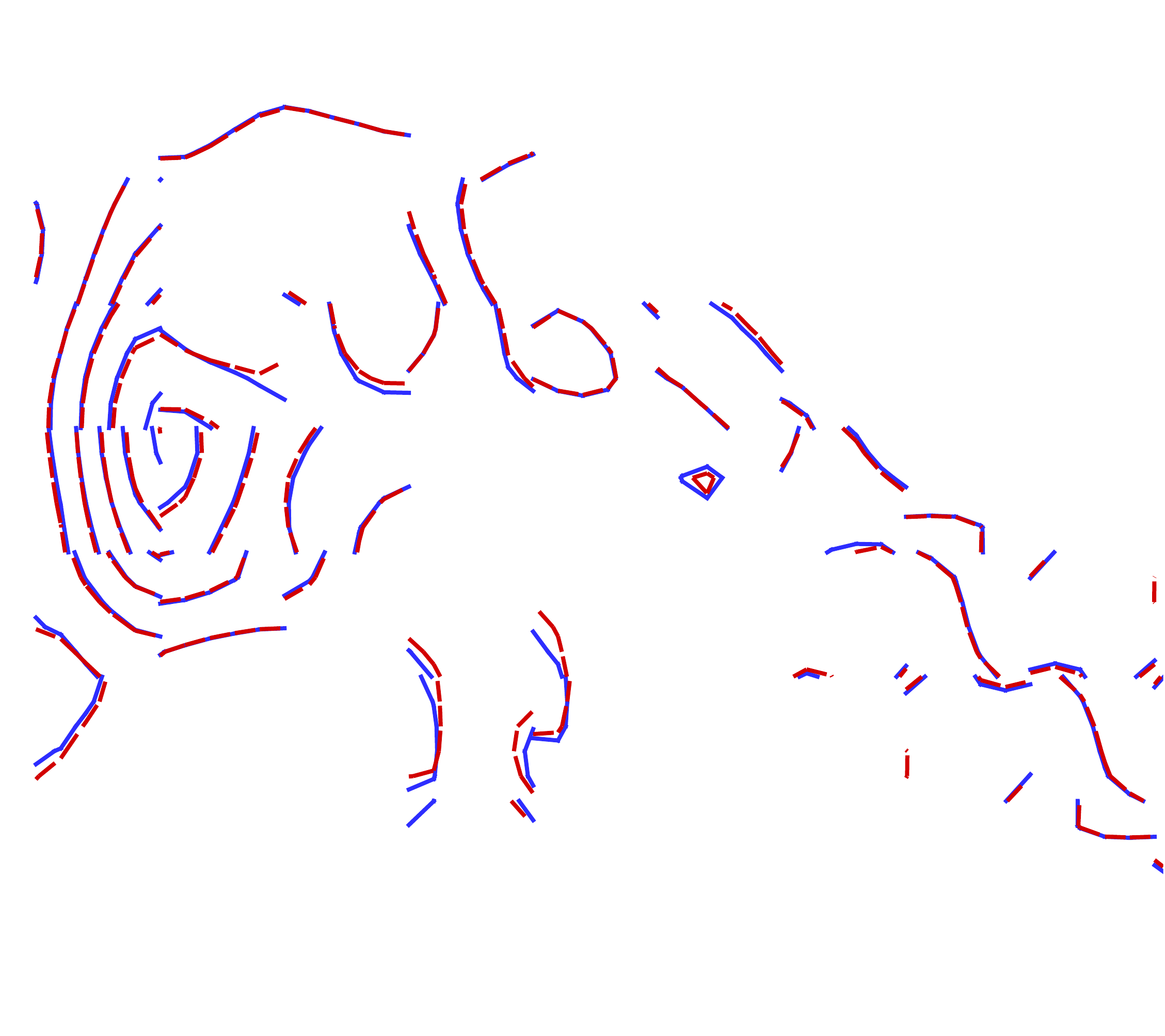}
				\par\end{centering}
			\centering{}\caption{Comparison of contours of surface normal vorticity on $y=\pi$ at $t=8$ in Taylor-Green
				vortex. Comparison between ERK2 (red dashed lines) and SDIRK2 with $\Delta t=0.1$ (blue solid lines).\label{fig:Comparison-of-surface} }
		\end{figure}

		The evolution of the Taylor-Green vortex (TGV) includes the break
		down of initially large vortices, transition to turbulence and decay
		of turbulence. With a very simple set up but a complex flow evolution,
		it is a very good test case for transitional and turbulent simulations.
		Initially, large vortices are placed in a cubic domain $[-\pi,\pi]^{3}$ with periodic
		boundary conditions. Flow conditions corresponding to $\text{Ma}=0.1$
		and $\text{Re}=1600$ are used. The expression
		of the initial flow field can be found in reference \citep{wiart_assessment_2014}.
		The value $2\mu\varepsilon/\rho$, where $\varepsilon$
		is the enstrophy, is a good estimate of the energy
		dissipation rate at the incompressible limit \citep{wiart_assessment_2014}.
		Fig.~\ref{fig:Different-resolutions} displays
		evolution curves of $2\mu\varepsilon/\rho$ obtained with $32^{3}$ meshes and using
		resolutions at different polynomial orders.
		Dealiasing via over-integration with $3(P+1)/2$ quadrature points in
		each spatial direction \citep{winters_comparative_2018} is used in
		the under-resolved simulations of TGV. The
		curves gradually tend to the DNS result \citep{wiart_assessment_2014} as the resolution increases,
		which shows the implicit solver can correctly resolve the main flow
		phenomena.  All
		these simulations are run using SDIRK2 with $\text{CFL}_{c}$ smaller than 2.5. The details of the implicit solver and their comparisons with ERK2 scheme are presented in Tab.~\ref{tab:Efficiency-comparison-between-2-1}. Although there is no grid stretching in this test case, the implicit solver can still be as efficient as, or more efficient than, the explicit solver. Fig.~\ref{fig:Different-temporal-methods} compares results
		calculated using SDIRK2 with different time steps with the result obtained
		with ERK2, which is very accurate in time because of the very small
		time step used ($\text{CFL}=0.08$). Three different time steps of $0.05$,
		$0.1$ and $0.2$ are used for SDIRK2. Although the corresponding temporal errors of SDIRK2
		should change by $16$ times for different time steps, all these
		results coincide quite well with the results of ERK2. This means the
		temporal error is very small for the implicit solver. Given the large differences for different spatial resolutions, the implicit
		solver with the current parameter choice is accurate enough in the
		sense that the error of the solution is dominated by the spatial error.
		Finally, Fig. \ref{fig:Comparison-of-surface} shows instantaneous vorticity distributions obtained using ERK2 and SDIRK2. They are in good agreement with each other, which further verify the accuracy of the implicit solver.

	\subsection{Turbulent flow over a circular cylinder at $\text{Re}=3900$} \label{3D_cylinder}

\begin{figure}
			\begin{centering}
				\includegraphics[width=0.75\textwidth]{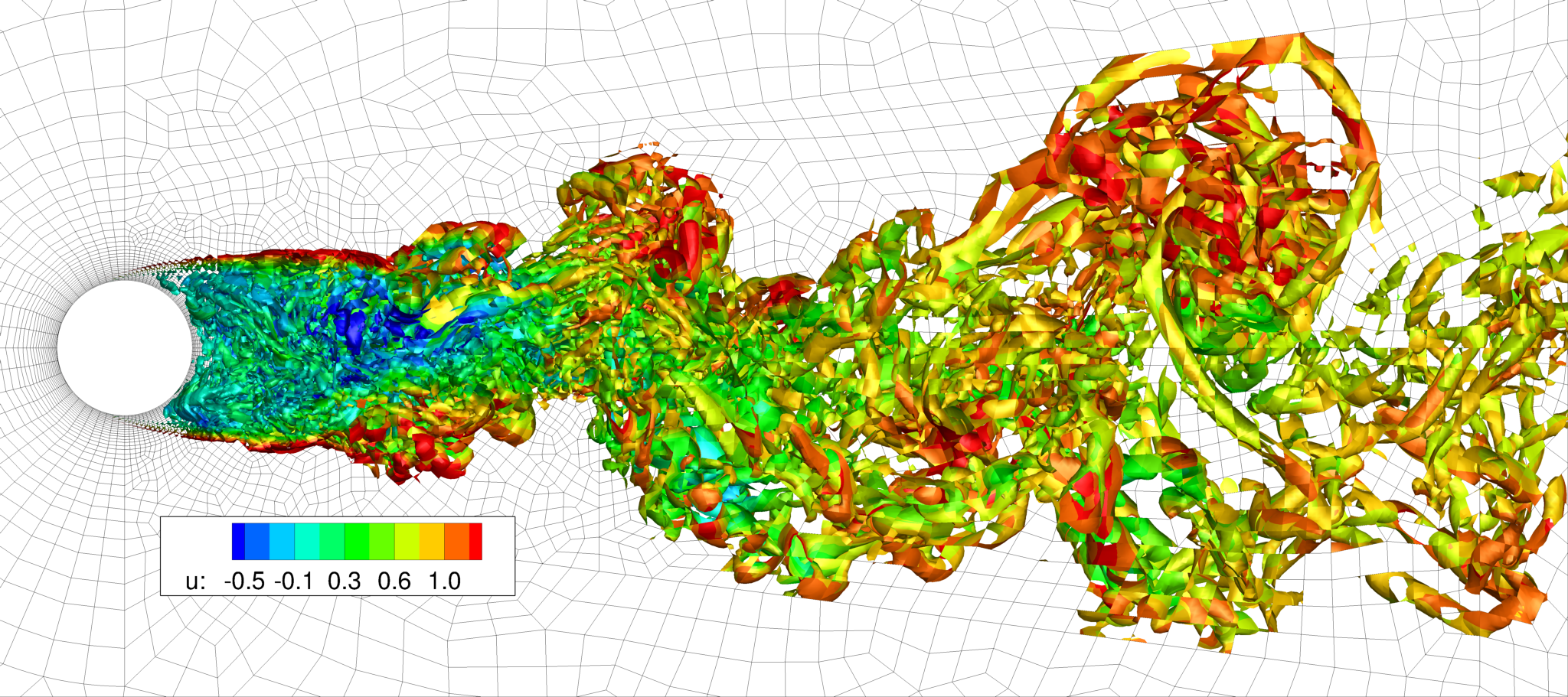}
				\par\end{centering}
			\centering{}\caption{Q criteria iso-surface ($Q=5$) and the mesh of turbulent
				flow over a circular cylinder at $\text{Re}=3900$\label{fig:Q-criteria-iso-surface}}
		\end{figure}

\begin{figure}
			\begin{centering}
				\includegraphics[width=0.5\textwidth]{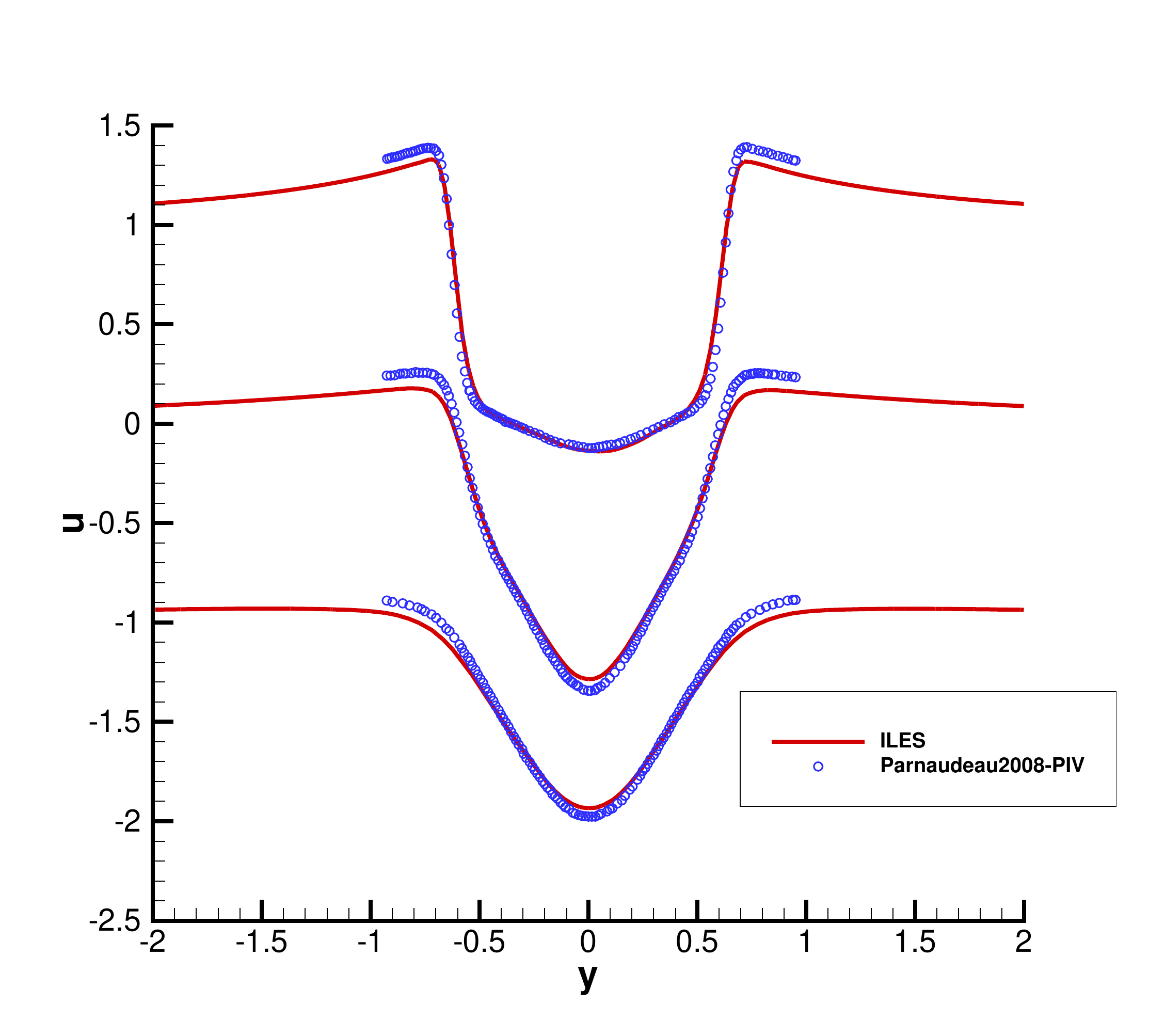}\includegraphics[width=0.5\textwidth]{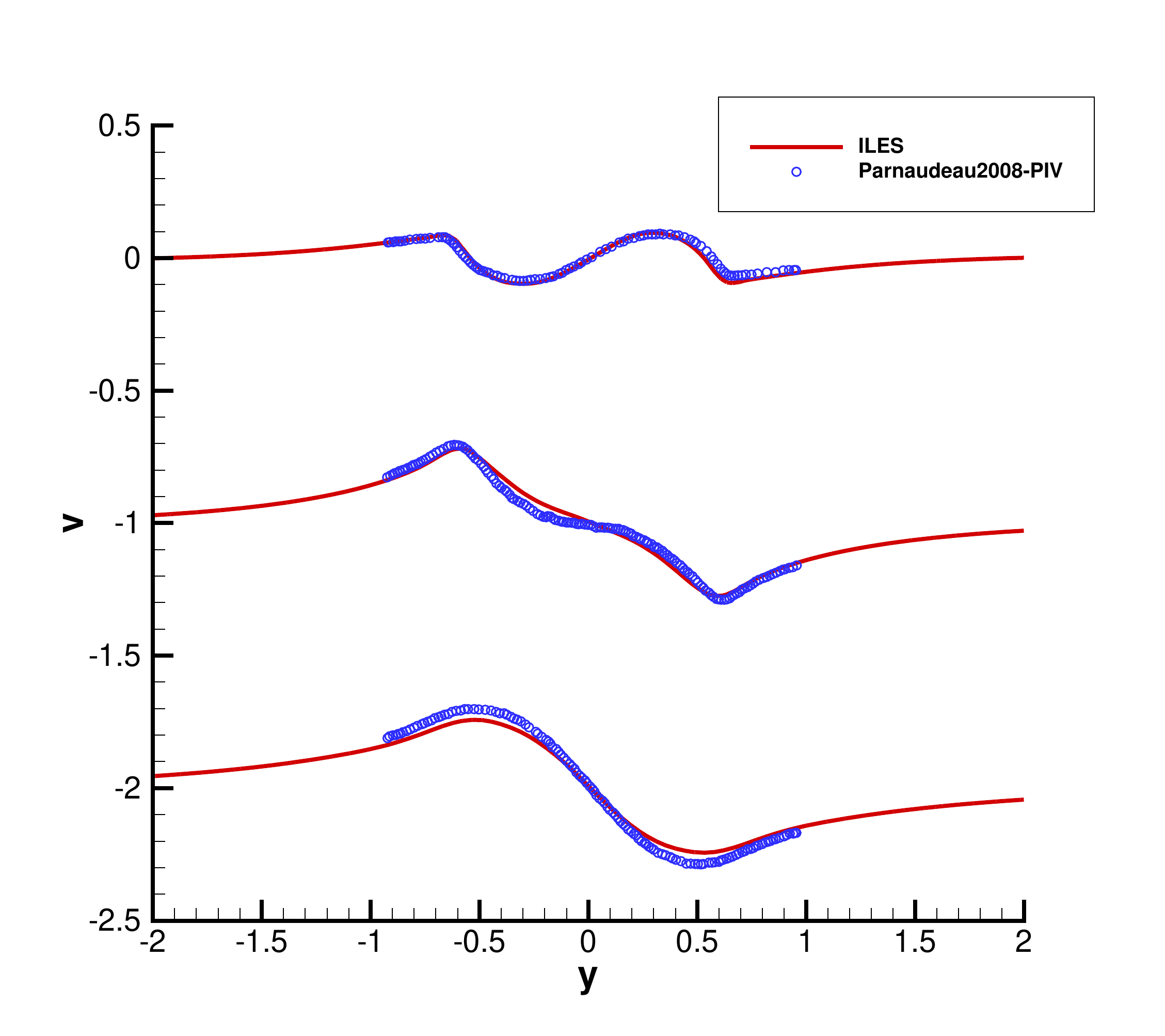}
				\par\end{centering}
			\centering{}\caption{Time averaged velocity distributions of turbulent flow over a circular cylinder
				at $\text{Re}=3900$\label{fig:Time-averaged-velocity}}
		\end{figure}

		The flow over a circular cylinder at $\text{Re}=3900$ is another
		widely used test case for turbulence simulations. Various experimental
		and numerical results can be found for this test case,
		e.g. \citep{parnaudeau_experimental_2008,kravchenko_numerical_2000}.
		Compared with the TGV, there are extra complexities because of
		grid stretching, the need for simulating the boundary layer, and the instability
		of the shear layer. The computational domain boundaries are located
		$40D$ away down stream and $20D$ away on other directions. Riemann
		invariant based boundary conditions are applied at the outer boundaries.
		The mesh near the cylinder is
		shown in Fig.~\ref{fig:Q-criteria-iso-surface}. The mesh consists of $123\,360$ curved
		unstructured hexahedral elements in total, which corresponds to
		about 3.3M DoFs for $P=2$. The smallest mesh size is approximately $0.006D$, which is located on the
		cylinder surface and is chosen based on the estimate advocated by
		reference \citep{kravchenko_numerical_2000}.
		As in the TGV test case, we use $P=2$ with over-integration. The computed vortex structure using $P=2$ is illustrated in Fig.~\ref{fig:Q-criteria-iso-surface},
		which shows the complex turbulent flow structures in the wake of the
		cylinder. The time-averaged velocity distributions in Fig.~\ref{fig:Time-averaged-velocity}
		show that the current result is in good agreement with the experimental
		results \citep{parnaudeau_experimental_2008}.
		
		As shown in Tab.~\ref{tab:Efficiency-comparison-3900CYLINDER}, a value $\text{CFL}=40$
		is employed in the simulations. The $\text{CFL}_{c}$ is around $2.5$
		at this CFL number, which ensures the temporal accuracy of the
		simulations. The corresponding time step is about 270 times larger
		than the explicit one with $\text{CFL}=0.15$. On average, the
		implicit solver converges using 3 Newton iterations per implicit RK stage and 3.0 GMRES
		iterations per Newton iterations. The implicit solver is 14.7 times faster than the explicit
		one. The $P=3$ case is also run for efficiency comparison and, as shown in Tab.~\ref{tab:Efficiency-comparison-3900CYLINDER}, a speed-up of 12.2 is achieved compared with the explicit solver. The seven stages third-order low-storage ERK scheme (ERK3-7) of Tab. A.15 of reference \citep{toulorge_optimal_2012} is also compared, which runs at a much larger CFL number ($\text{CFL}=0.6$). The implicit solver is 11.5 times faster than the ERK3-7 for the simulation with $P=2$.

		The influence of the Newton iteration tolerance $\alpha$ is also studied here. With $\alpha=10^{-4}$, $\alpha=10^{-5}$, $\alpha=10^{-6}$ and $\alpha=10^{-7}$, the CPU time normalized by CPU time with $\alpha=10^{-3}$ is 1.0, 1.3, 1.6 and 1.7, respectively. Therefore decreasing $\alpha$ will not seriously slow down the simulations.


		\begin{table}[H]
			\begin{centering}
			\begin{tabular}{|c||c|c||c|c|}
			\hline 
			\textbf{\scriptsize{}Polynomial order} & \multicolumn{2}{c||}{$P=2$} & \multicolumn{2}{c|}{$P=3$}\tabularnewline
			\hline 
			\textbf{\scriptsize{}Method} & \textbf{\scriptsize{}ERK2} & \textbf{\scriptsize{}SDIRK2} & \textbf{\scriptsize{}ERK2} & \textbf{\scriptsize{}SDIRK2}\tabularnewline
			\hline 
			\hline 
			\textbf{\scriptsize{}$\Delta t$} & {\scriptsize{}$3.5\times10^{-5}$} & {\scriptsize{}$1.0\times10^{-2}$} & {\scriptsize{}$1.56\times10^{-5}$} & {\scriptsize{}$5.0\times10^{-3}$}\tabularnewline
			\hline 
			{\scriptsize{}$\text{CFL}$} & {\scriptsize{}0.15} & {\scriptsize{}40} & {\scriptsize{}0.15} & {\scriptsize{}48}\tabularnewline
			\hline 
			{\scriptsize{}$\text{CFL}_{c}$} & {\scriptsize{}$9.0\times10^{-3}$} & {\scriptsize{}2.5} & {\scriptsize{}$6.0\times10^{-3}$} & {\scriptsize{}1.7}\tabularnewline
			\hline 
			\textbf{\scriptsize{}Newton iterations} & \multirow{2}{*}{{\scriptsize{}\textbackslash{}}} & \multirow{2}{*}{{\scriptsize{}3}} & \multirow{2}{*}{{\scriptsize{}\textbackslash{}}} & \multirow{2}{*}{{\scriptsize{}3}}\tabularnewline
			\textbf{\scriptsize{}per RK stage} &  &  &  & \tabularnewline
			\hline 
			\textbf{\scriptsize{}GMRES iterations} & \multirow{2}{*}{{\scriptsize{}\textbackslash{}}} & \multirow{2}{*}{{\scriptsize{}3.0}} & \multirow{2}{*}{{\scriptsize{}\textbackslash{}}} & \multirow{2}{*}{{\scriptsize{}3.7}}\tabularnewline
			\textbf{\scriptsize{}per Newton iteration} &  &  &  & \tabularnewline
			\hline 
			\textbf{\scriptsize{}Speed-up} & {\scriptsize{}1.0} & {\scriptsize{}14.7} & {\scriptsize{}1.0} & {\scriptsize{}12.2}\tabularnewline
			\hline 
			\end{tabular}
			\par\end{centering}
			\caption{Efficiency comparison between explicit and implicit time integration
			schemes of turbulent circular cylinder at $\text{Re}=3900$ \label{tab:Efficiency-comparison-3900CYLINDER}}
		\end{table}

	\subsection{Shock wave boundary-layer interaction}\label{SWBLI}

\begin{figure}
			\begin{centering}
				\subfloat[Mach number distribution\label{fig:Mach-number-distribution}]{\begin{centering}
						\includegraphics[width=0.5\textwidth]{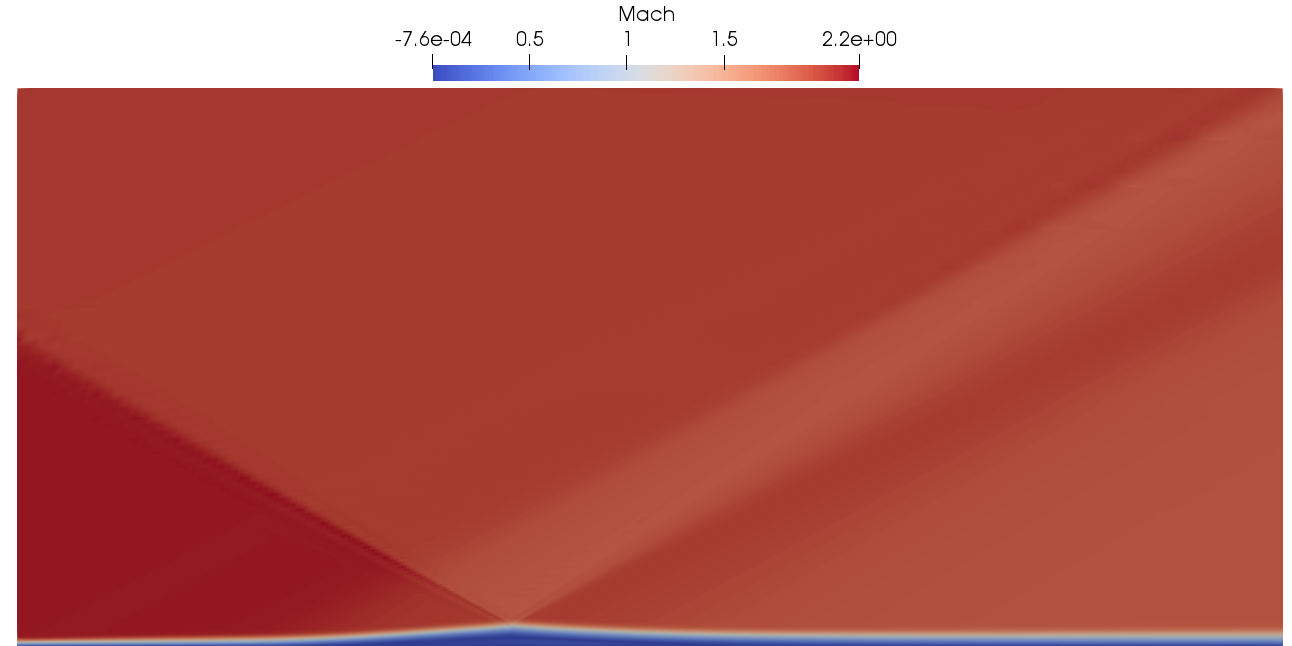}
						\par\end{centering}
					\begin{centering}
						\par\end{centering}
				}\subfloat[Skin friction coefficient distribution. Blue line: SDIRK-2; Red line:
					ERK2; triangles: simulation from \citep{boin_3d_2006}; dotted line:
					empirical solution by \citep{eckert_engineering_1955}\label{fig:Skin-friction-coefficient-1}]{\begin{centering}
						\includegraphics[width=0.5\textwidth]{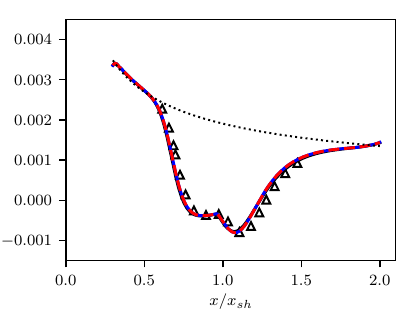}
						\par\end{centering}
					\begin{centering}
						\par\end{centering}
				}
				\par\end{centering}
			\centering{}\caption{Mach number and skin friction coefficient distributions in shock wave
				boundary-layer interaction }
		\end{figure}

		Finally, the shock capturing ability of the implicit solver is tested
		using the shock wave boundary-layer interaction (SWBLI) problem \citep{boin_3d_2006}.
		The computational domain and Mach number distribution are displayed in Fig.~\ref{fig:Mach-number-distribution}.
		The viscous wall starts from $x=0$ and the shock impingement position
		is $x_{sh}$. The computational domain starts from $x=0.3x_{sh}$ where
		the analytical compressible boundary layer solution \citep{white_viscous_1991}
		is imposed. Rankine-Hugoniot relations are added to impose inflow boundary conditions consistent with the shock.
		Flow conditions corresponding to $\text{Ma}=2.15$, $\text{Re}_{x_{sh}}=10^{5}$
		and shock impingement angle of $30.8$ are used. The simulations run
		with $119\times39$ quadrilateral elements and $P=3$. The skin friction coefficient
		distributions are shown in Fig.~\ref{fig:Skin-friction-coefficient-1}.
		The results of the explicit and implicit solvers coincide and are in good agreement with the simulation results
		presented in reference \citep{boin_3d_2006}. The efficiency of the
		solvers is compared in Tab.~\ref{tab:Efficiency-comparison-SWBLI}.
		Compared with the explicit solver, larger speed-up can be achieved when the CFL number of the implicit solver increases from 2.5 to 100. However, further increasing the CFL number to 1000, the implicit solver will have difficulty converging. The optimal numbers of preconditioning iterations are also given in Tab.~\ref{tab:Efficiency-comparison-SWBLI}. With a smaller time step, the system is less stiff and the optimal preconditioning number is smaller.  Since the
		test case is regarded as a steady-state problem, a large CFL number
		can be used without influencing the accuracy. A speed-up of 22.8 is
		achieved with CFL=100. 

		\begin{table}[H]
			\begin{centering}
			\begin{tabular}{|c||c|c|c|c|}
			\hline 
			\textbf{\scriptsize{}Polynomial order} & \textbf{\scriptsize{}ERK2} & \multicolumn{3}{c|}{\textbf{\scriptsize{}SDIRK2}}\tabularnewline
			\hline 
			\hline 
			{\scriptsize{}$\text{CFL}$} & {\scriptsize{}0.08} & {\scriptsize{}2.5} & {\scriptsize{}5.0} & {\scriptsize{}100.0}\tabularnewline
			\hline 
			\textbf{\scriptsize{}$\Delta t$} & {\scriptsize{}$1.3\times10^{-4}$} & {\scriptsize{}$4.0\times10^{-3}$} & {\scriptsize{}$8.0\times10^{-3}$} & {\scriptsize{}$1.6\times10^{-2}$}\tabularnewline
			\hline 
			$J$ in BRJ & \textbackslash{} & {\scriptsize{}3} & {\scriptsize{}5} & {\scriptsize{}7}\tabularnewline
			\hline 
			\textbf{\scriptsize{}Newton iterations} & \multirow{2}{*}{{\scriptsize{}\textbackslash{}}} & \multirow{2}{*}{{\scriptsize{}2.0}} & \multirow{2}{*}{{\scriptsize{}2.0}} & \multirow{2}{*}{{\scriptsize{}3.0}}\tabularnewline
			\textbf{\scriptsize{}per RK stage} &  &  &  & \tabularnewline
			\hline 
			\textbf{\scriptsize{}GMRES iterations} & \multirow{2}{*}{{\scriptsize{}\textbackslash{}}} & \multirow{2}{*}{{\scriptsize{}3.0}} & \multirow{2}{*}{{\scriptsize{}3.2}} & \multirow{2}{*}{{\scriptsize{}8.9}}\tabularnewline
			\textbf{\scriptsize{}per Newton iteration} &  &  &  & \tabularnewline
			\hline 
			\textbf{\scriptsize{}Speed-up} & {\scriptsize{}1.0} & {\scriptsize{}2.7} & {\scriptsize{}4.2} & {\scriptsize{}22.8}\tabularnewline
			\hline 
			\end{tabular}
			\par\end{centering}
			\caption{Efficiency comparison between explicit and implicit time integration
			schemes of shock wave boundary-layer interaction \label{tab:Efficiency-comparison-SWBLI}}
		\end{table}

\section{Discussion and future work}

	An implicit flow solver for the compressible Navier-Stokes equations has been developed in Nektar++ to improve the simulation efficiency using singly diagonally implicit Runge-Kutta and Jacobian-free Newton 	Krylov methods. To improve the efficiency, an efficient block relaxed Jacobi (BRJ) preconditioner is proposed, the computational cost and memory consumption of which are reduced by approximating the element boundary integrations, using single data precision and using a matrix-free implementation of the off diagonal terms. Performance analysis shows that the proposed BRJ preconditioner is low in computational cost at least for $P=2$ and $P=3$, which enables the solver to use more preconditioning iterations. The memory consumption of BRJ is significantly smaller than an ILU preconditioner. The software design and implementation details have
	been described with emphasis on providing a general pattern for the development
	of implicit solvers and for reducing
	the memory footprint. The solver capabilities were summarized and
	demonstrated using a variety of verification test cases. In the isentropic vortex
	convection, manufactured compressible Poiseuille flow and laminar
	flow over a circular cylinder test cases, the results achieved the designed
	convergent order of accuracy, thus verifying the implementations of
	the advection, diffusion and temporal discretization methods.
	The results of the TGV test case showed that the implicit
	solver achieves good temporal accuracy in the sense that the temporal
	error is much smaller than the spatial error. The implicit
	solver produced accurate predictions of turbulent
	flow over a circular cylinder at $\text{Re}=3900$ and shock wave boundary-layer
	interaction.  The computational efficiency of the implicit solver is
	more than 10 times that of the explicit solver in both steady
	and unsteady subsonic simulations as well as steady supersonic simulations.
	The implicit solver is recommended for simulations with large grid
	stretching and/or low Mach number.

	Further studies will focus on the development and implementation of efficient
	preconditioning methods. Moreover, the performance of the implicit solver for supersonic
	simulations has only been preliminarily investigated and will require to be studied
	thoroughly.

\section*{Acknowledgements}

	The development the implicit solver in Nektar++ has
	been supported by EPSRC grant (EP/R029423/1) and UK  Turbulence
	Consortium grant (EP/R029326/1). We would like to gratefully
	acknowledge access to the computing facilities provided by the Imperial College
	Research Computing Service (DOI: 10.14469/hpc/2232).
	Zhen-Guo Yan acknowledges support from the National Natural
	Science Foundation of China (Grant No. 11902344).

\section*{Appendix A: Butcher tableaus of Runge-Kutta schemes}

	The coefficients of Runge-Kutta schemes are given using Butcher tableaus.
	The coefficients of the explicit second-order Runge-Kutta scheme (ERK2), the second-order singly diagonally implicit Runge-Kutta scheme (SDIRK2), the third-order singly diagonally implicit Runge-Kutta scheme (SDIRK3), the four-order singly diagonally implicit Runge-Kutta scheme (SDIRK4),  are given in
	Tab.~\ref{tab:RK-coeff}.

	\begin{table}
		\begin{centering}
			\subfloat[ERK2 \label{tab:rk2}]{\begin{centering}
					\begin{tabular}{c|cc}
						0.0 &     & \tabularnewline
						1.0 & 1.0 & \tabularnewline
						\hline
							& 0.5 & 0.5\tabularnewline
					\end{tabular}
					\par\end{centering}
				\centering{}}
			\subfloat[SDIRK2 with $\lambda=1-\sqrt{2}/2$. \label{tab:Second-order-diagonally-implicit}]{\begin{centering}
					\begin{tabular}{c|cc}
						$\lambda$ & $\lambda$     & \tabularnewline
						1.0       & 1.0-$\lambda$ & $\lambda$\tabularnewline
						\hline
									& 1.0-$\lambda$ & $\lambda$\tabularnewline
					\end{tabular}
					\par\end{centering}
				\centering{}}

			\subfloat[SDIRK3 with $\lambda=0.4358665215$]{\begin{centering}
				\begin{tabular}{c|ccc}
				$\lambda$ & $\lambda$ &  & \tabularnewline
				$\frac{1+\lambda}{2}$ & $\frac{1-\lambda}{2}$ & $\lambda$ & \tabularnewline
				1 & $\frac{-6\lambda^{2}+16\lambda-1}{4}$ & $\frac{6\lambda^{2}-20\lambda+5}{4}$ & $\lambda$\tabularnewline
				\hline 
					& $\frac{-6\lambda^{2}+16\lambda-1}{4}$ & $\frac{6\lambda^{2}-20\lambda+5}{4}$ & $\lambda$\tabularnewline
				\end{tabular}
				\par\end{centering}
				\centering{}}

				\subfloat[SDIRK4 with 6 stages \citep{kennedy_diagonally_2016} ]{\begin{centering}
					\begin{tabular}{c|cccccc}
					0 & 0 &  &  &  &  & \tabularnewline
					$\frac{1}{2}$ & $a^{11}$ & $\frac{1}{4}$ &  &  &  & \tabularnewline
					$\frac{2-\sqrt{2}}{2}$ & $a^{21}$ & $\frac{1-\sqrt{2}}{8}$ & $\frac{1}{4}$ &  &  & \tabularnewline
					$\frac{5}{8}$ & $a^{31}$ & $\frac{5-7\sqrt{2}}{64}$ & $\frac{7(1+\sqrt{2})}{32}$ & $\frac{1}{4}$ &  & \tabularnewline
					$\frac{26}{25}$ & $a^{41}$ & $\frac{-13796-54539\sqrt{2}}{125000}$ & $\frac{506605+132109\sqrt{2}}{437500}$ & $\frac{166(-97+376\sqrt{2})}{109375}$ & $\frac{1}{4}$ & \tabularnewline
					1 & $a^{51}$ & $\frac{1181-987\sqrt{2}}{13782}$ & $\frac{47(-267+1783\sqrt{2})}{273343}$ & $\frac{-16(-22922+3525\sqrt{2})}{571953}$ & $\frac{-15625(97+376\sqrt{2})}{90749876}$ & $\frac{1}{4}$\tabularnewline
					\hline 
					 & $a^{50}$ & $a^{51}$ & $a^{52}$ & $a^{53}$ & $a^{54}$ & $a^{55}$\tabularnewline
					\end{tabular}
					\par\end{centering}
					\centering{}}
					
			\par\end{centering}
		\caption{Butcher tableaus of Runge-Kutta schemes} \label{tab:RK-coeff}

	\end{table}

\section*{Appendix B: Forcing terms of manufactured compressible Poiseuille
			flow}

	The analytical solution of the manufactured compressible Poiseuille
	flow has already been given in Eq.~\eqref{eq:MMS-SOLUTION}. The corresponding
	forcing terms are given by
	\begin{equation}
		f=\left(\begin{array}{c}
				0                                                \\
				\frac{dp}{dx}\theta\left(L^{2}-6yL+6y^{2}\right) \\
				0                                                \\
				f_{4}
			\end{array}\right),\label{eq:forcing-MMS}
	\end{equation}
	where
	\begin{equation}
		\begin{aligned}f_{4}= & \frac{1}{2\mu}\frac{dp}{dx}^{2}(y(L-y)\left(1-\theta(L^{2}-6yL+6y^{2})-\gamma/(\gamma-1)\right) \\
					& -\frac{1}{4\mu}\frac{dp}{dx}^{2}(L-2y)^{2}(2\theta yL-2\theta y^{2}+1)^{2}.
		\end{aligned}
		\label{eq:FORCING-MMS1}
	\end{equation}

	\bibliographystyle{spmpsci_esr}
	\bibliography{CAMWA-high-order-issue-references}

\begin{thebibliography}{10}
\providecommand{\url}[1]{{#1}}
\providecommand{\urlprefix}{URL }
\expandafter\ifx\csname urlstyle\endcsname\relax
  \providecommand{\doi}[1]{DOI~\discretionary{}{}{}#1}\else
  \providecommand{\doi}{DOI~\discretionary{}{}{}\begingroup
  \urlstyle{rm}\Url}\fi

\bibitem{arnold_unified_2002}
Arnold, D., Brezzi, F., Cockburn, B., Marini, L.: Unified analysis of
  discontinuous {G}alerkin methods for elliptic problems.
\newblock SIAM Journal on Numerical Analysis \textbf{39}(5), 1749--1779 (2002)

\bibitem{bastian_matrix-free_2019}
Bastian, P., M\"uller, E.H., M\"uthing, S., Piatkowski, M.: Matrix-free
  multigrid block-preconditioners for higher order discontinuous {Galerkin}
  discretisations.
\newblock Journal of Computational Physics \textbf{394}, 417--439 (2019)

\bibitem{bijl_implicit_2002}
Bijl, H., Carpenter, M.H., Vatsa, V.N., Kennedy, C.A.: Implicit time
  integration schemes for the unsteady compressible
  {Navier}{\textendash}{Stokes} equations: Laminar flow.
\newblock Journal of Computational Physics \textbf{179}(1), 313--329 (2002)

\bibitem{boin_3d_2006}
Boin, J.P., Robinet, J.C., Corre, C., Deniau, H.: 3{D} steady and unsteady
  bifurcations in a shock-wave/laminar boundary layer interaction: A numerical
  study.
\newblock Theoretical and Computational Fluid Dynamics \textbf{20}(3), 163--180
  (2006)

\bibitem{brune_composing_2015}
Brune, P.R., Knepley, M.G., Smith, B.F., Tu, X.: Composing scalable nonlinear
  algebraic solvers.
\newblock SIAM Review \textbf{57}(4), 535--565 (2015)

\bibitem{Cantwell2015}
Cantwell, C.D., Moxey, D., Comerford, A., Bolis, A., Rocco, G., Mengaldo, G.,
  {De Grazia}, D., Yakovlev, S., Lombard, J.E., Ekelschot, D., Jordi, B., Xu,
  H., Mohamied, Y., Eskilsson, C., Nelson, B., Vos, P., Biotto, C., Kirby,
  R.M., Sherwin, S.J.: Nektar++: {A}n open-source spectral/hp element
  framework.
\newblock Computer Physics Communications \textbf{192}, 205--219 (2015)

\bibitem{cantwell_h_2011}
Cantwell, C.D., Sherwin, S.J., Kirby, R.M., Kelly, P.H.J.: From h to p
  efficiently: {Strategy} selection for operator evaluation on hexahedral and
  tetrahedral elements.
\newblock Computers \& Fluids \textbf{43}(1), 23--28 (2011)

\bibitem{cheng_direct_2016}
Cheng, J., Yang, X., Liu, X., Liu, T., Luo, H.: A direct discontinuous
  {G}alerkin method for the compressible {N}avier-{S}tokes equations on
  arbitrary grids.
\newblock Journal of Computational Physics \textbf{327}, 484--502 (2016)

\bibitem{chudanov_validation_2014}
Chudanov, V., Aksenova, A., Goreinov, S., Makarevich, A., Pervichko, V.:
  Validation of a new method for solving of {CFD} problems in nuclear
  engineering using petascale {HPC}.
\newblock pp. V004T10A009--V004T10A009. American Society of Mechanical
  Engineers (2014)

\bibitem{cockburn_local_1998}
Cockburn, B., Shu, C.: The local discontinuous {G}alerkin method for
  time-dependent convection-diffusion systems.
\newblock SIAM Journal on Numerical Analysis \textbf{35}(6), 2440--2463 (1998)

\bibitem{diosady_scalable_2019}
Diosady, L.T., Murman, S.M.: Scalable tensor-product preconditioners for
  high-order finite-element methods: {Scalar} equations.
\newblock Journal of Computational Physics \textbf{394}, 759--776 (2019)

\bibitem{ducros_large-eddy_1999}
Ducros, F., Ferrand, V., Nicoud, F., Weber, C., Darracq, D., Gacherieu, C.,
  Poinsot, T.: Large-eddy simulation of the shock/turbulence interaction.
\newblock Journal of Computational Physics \textbf{152}(2), 517--549 (1999)

\bibitem{eckert_engineering_1955}
Eckert, E.: Engineering relations for friction and heat transfer to surfaces in
  high velocity flow.
\newblock Journal of the Aeronautical Sciences \textbf{22}(8), 585--587 (1955)

\bibitem{ezertas_performances_2009}
Ezertas, A., Eyi, S.: Performances of numerical and analytical jacobians in
  flow and sensitivity analysis.
\newblock In: 19th {AIAA} {Computational} {Fluid} {Dynamics}. American
  Institute of Aeronautics and Astronautics, San Antonio, Texas (2009)

\bibitem{Franciolini2017}
Franciolini, M., Crivellini, A., Nigro, A.: On the efficiency of a matrix-free
  linearly implicit time integration strategy for high-order discontinuous
  {G}alerkin solutions of incompressible turbulent flows.
\newblock Computers and Fluids \textbf{159}, 276--294 (2017)

\bibitem{hartmann_symmetric_2005}
Hartmann, R., Houston, P.: Symmetric interior penalty {DG} methods for the
  compressible {N}avier-{S}tokes equations {I}: {M}ethod formulation.
\newblock International Journal of Numerical Analysis and Modeling
  \textbf{3}(1), 1--20 (2005)

\bibitem{hartmann_optimal_2008}
Hartmann, R., Houston, P.: An optimal order interior penalty discontinuous
  {Galerkin} discretization of the compressible {Navier}{\textendash}{Stokes}
  equations.
\newblock Journal of Computational Physics \textbf{227}(22), 9670--9685 (2008)

\bibitem{hesthaven_nodal_2007}
Hesthaven, J.S., Warburton, T.: Nodal Discontinuous {G}alerkin Methods:
  Algorithms, Analysis, and Applications.
\newblock Springer Science \& Business Media (2007)

\bibitem{hillewaert_development_2013}
Hillewaert, K.: Development of the discontinuous {G}alerkin method for
  high-resolution, large scale {CFD} and acoustics in industrial geometries.
\newblock Ph.D. thesis, Universit{\'e} Catholique de Louvain (2013)

\bibitem{karniadakis_spectral/hp_2013}
Karniadakis, G., Sherwin, S.: Spectral/hp {Element} {Methods} for
  {Computational} {Fluid} {Dynamics}, second edn.
\newblock Oxford Science Publications (2013)

\bibitem{kennedy_diagonally_2016}
Kennedy, C.A., Carpenter, M.H.: Diagonally implicit {R}unge-{K}utta methods for
  ordinary differential equations. {A} review.
\newblock {NASA} report TM-2016-219173 (2016)

\bibitem{knoll_jacobian-free_2004}
Knoll, D.A., Keyes, D.E.: Jacobian-free {N}ewton-{K}rylov methods: {A} survey
  of approaches and applications.
\newblock Journal of Computational Physics \textbf{193}(2), 357--397 (2004)

\bibitem{kravchenko_numerical_2000}
Kravchenko, A.G., Moin, P.: Numerical studies of flow over a circular cylinder
  at {Re}{$_D$}=3900.
\newblock Physics of Fluids \textbf{12}(2), 403--417 (2000)

\bibitem{Mengaldo2015}
Mengaldo, G.: Discontinuous spectral/hp element methods: development, analysis
  and applications to compressible flows.
\newblock Ph.D. thesis, Imperial College London (2015)

\bibitem{mengaldo_connections_2016}
Mengaldo, G., De~Grazia, D., Vincent, P.E., Sherwin, S.J.: On the connections
  between discontinuous {G}alerkin and flux reconstruction schemes: {E}xtension
  to curvilinear meshes.
\newblock Journal of Scientific Computing \textbf{67}(3), 1272--1292 (2016)

\bibitem{moxey_nektar++:_2019}
Moxey, D., Cantwell, C.D., Bao, Y., Cassinelli, A., Castiglioni, G., Chun, S.,
  Juda, E., Kazemi, E., Lackhove, K., Marcon, J., Mengaldo, G., Serson, D.,
  Turner, M., Xu, H., Peir{\'o}, J., Kirby, R.M., Sherwin, S.J.: Nektar++:
  {E}nhancing the capability and application of high-fidelity spectral/$hp$
  element methods.
\newblock Computer Physics Communications  (2019)

\bibitem{oliver_multigrid_2004}
Oliver, T.A.: Multigrid solution for high-order discontinuous {Galerkin}
  discretizations of the compressible {Navier}-{Stokes} equations.
\newblock Thesis, Massachusetts Institute of Technology (2004)

\bibitem{orszag_spectral_1980}
Orszag, S.A.: Spectral methods for problems in complex geometries.
\newblock Journal of Computational Physics \textbf{37}(1), 70--92 (1980)

\bibitem{parnaudeau_experimental_2008}
Parnaudeau, P., Carlier, J., Heitz, D., Lamballais, E.: Experimental and
  numerical studies of the flow over a circular cylinder at {Reynolds} number
  3900.
\newblock Physics of Fluids \textbf{20}(8), 085101 (2008)

\bibitem{pazner_approximate_2018}
Pazner, W., Persson, P.O.: Approximate tensor-product preconditioners for very
  high order discontinuous {Galerkin} methods.
\newblock Journal of Computational Physics \textbf{354}, 344--369 (2018)

\bibitem{peraire_compact_2008}
Peraire, J., Persson, P.: The compact discontinuous {G}alerkin ({CDG}) method
  for elliptic problems.
\newblock SIAM Journal on Scientific Computing \textbf{30}(4), 1806--1824
  (2008)

\bibitem{persson_sub-cell_2006}
Persson, P.O., Peraire, J.: Sub-cell shock capturing for discontinuous
  {G}alerkin methods.
\newblock In: 44th {AIAA} {Aerospace} {Sciences} {Meeting} and {Exhibit}. AIAA
  2006-112, Reno, Nevada (2006)

\bibitem{peterson_overview_2018}
Peterson, J.W., Lindsay, A.D., Kong, F.: Overview of the incompressible
  {Navier}{\textendash}{Stokes} simulation capabilities in the {MOOSE}
  framework.
\newblock Advances in Engineering Software \textbf{119}, 68--92 (2018)

\bibitem{roy_review_2005}
Roy, C.J.: Review of code and solution verification procedures for
  computational simulation.
\newblock Journal of Computational Physics \textbf{205}(1), 131--156 (2005)

\bibitem{saad_gmres:_1986}
Saad, Y., Schultz, M.H.: {GMRES}: {A} generalized minimal residual algorithm
  for solving nonsymmetric linear systems.
\newblock SIAM Journal on Scientific and Statistical Computing \textbf{7}(3),
  856--869 (1986)

\bibitem{toro_riemann_2009}
Toro, E.F.: Riemann Solvers and Numerical Methods for Fluid Dynamics, third
  edn.
\newblock Springer (2009)

\bibitem{toulorge_optimal_2012}
Toulorge, T., Desmet, W.: Optimal {Runge}-{Kutta} schemes for discontinuous
  {Galerkin} space discretizations applied to wave propagation problems.
\newblock Journal of Computational Physics \textbf{231}(4), 2067--2091 (2012)

\bibitem{vanden_comparison_1996}
Vanden, K.J., Orkwis, P.D.: Comparison of numerical and analytical {Jacobians}.
\newblock AIAA Journal \textbf{34}(6), 1125--1129 (1996)

\bibitem{vandenhoeck_implicit_2019}
Vandenhoeck, R., Lani, A.: Implicit high-order flux reconstruction solver for
  high-speed compressible flows.
\newblock Computer Physics Communications \textbf{242}, 1--24 (2019)

\bibitem{Vos2011}
Vos, P.E.J., Eskilsson, C., Bolis, A., Chun, S., Robert, M., Sherwin, S.J.: A
  generic framework for time-stepping partial differential equations ({PDE}s):
  {G}eneral linear methods, object-oriented implementation and application to
  fluid problems.
\newblock International Journal of Computational Fluid Dynamics \textbf{25}(3),
  107--125 (2011)

\bibitem{white_viscous_1991}
White, F.M.: Viscous fluid flow, second edn.
\newblock McGraw-Hill, New York (1991)

\bibitem{wiart_assessment_2014}
Wiart, C.C.d., Hillewaert, K., Duponcheel, M., Winckelmans, G.: Assessment of a
  discontinuous {Galerkin} method for the simulation of vortical flows at high
  {Reynolds} number.
\newblock International Journal for Numerical Methods in Fluids \textbf{74}(7),
  469--493 (2014)

\bibitem{williams_roofline:_2009}
Williams, S., Waterman, A., Patterson, D.: Roofline: {An} insightful visual
  performance model for multicore architectures.
\newblock Commun. ACM \textbf{52}(4), 65--76 (2009)

\bibitem{winters_comparative_2018}
Winters, A.R., Moura, R.C., Mengaldo, G., Gassner, G.J., Walch, S., Peir\'o,
  J., Sherwin, S.J.: A comparative study on polynomial dealiasing and split
  form discontinuous {Galerkin} schemes for under-resolved turbulence
  computations.
\newblock Journal of Computational Physics \textbf{372}, 1--21 (2018)

\bibitem{xiaoquan_robust_2019}
Xiaoquan, Y., Cheng, J., Luo, H., Zhao, Q.: Robust implicit direct
  discontinuous galerkin method for simulating the compressible turbulent
  flows.
\newblock AIAA Journal \textbf{57}(3), 1113--1132 (2019)

\end{thebibliography}

\end{document}